\documentclass[12pt]{article}
\usepackage{a4wide}
\usepackage{amssymb}
\usepackage{graphicx}
\usepackage[center]{subfigure}
\newcommand{\ba}{\begin{eqnarray}}
\newcommand{\ea}{\end{eqnarray}}
\newcommand{\be}{\begin{equation}}
\newcommand{\ee}{\end{equation}}

\begin{document}
\begin{titlepage}
\begin{flushright}
TTP04-20\\
SI-HEP-2004-09\\
\end{flushright}
\vfill
\begin{center}
{\Large\bf Modeling the Pion and Kaon Form Factors in the Timelike Region}
\\[2cm]
{\large\bf Christine Bruch~$^{a)}$, Alexander Khodjamirian~$^{b)}$, 
Johann~H.~K\"uhn~$^{a)}$ }\\[0.5cm]

{\it $^a)$Institut f\"ur Theoretische Teilchenphysik, Universit\"at
Karlsruhe,\\  D-76128 Karlsruhe, Germany}\\[0.2cm]
{\it $^b)$ Theoretische Physik 1, Fachbereich Physik, Universit\"at Siegen,\\ 
D-57068 Siegen, Germany}\\

\end{center}
\vfill
\begin{abstract}
New, accurate measurements of the pion and kaon electromagnetic form factors 
are expected in the near future from experiments at electron-positron
colliders,
using the radiative return method. We construct a model for the timelike pion 
electromagnetic form factor, that is valid also at momentum transfers 
far above the $\rho$ resonance. The ansatz  
is based on vector dominance and includes a pattern of radial excitations 
expected from dual resonance models.
The form factor is fitted to the existing data in the timelike region, continued  
to the spacelike region and compared with  
the measurements there and with the QCD predictions.
Furthermore, the model is extended to the kaon 
electromagnetic form factor. Using isospin
and SU(3)-flavour symmetry relations 
we extract the isospin-one contribution and predict the kaon weak form factor 
accessible in semileptonic $\tau$ decays.   
\end{abstract}
\vfill

\end{titlepage}

\section{Introduction}

The pion electromagnetic (e.m.) form factor $F_\pi(s)$,
one of the traditional study objects in hadron physics, 
nowadays plays an essential  role for the 
precise determination of electroweak observables. 
An accurate knowledge of $F_\pi(s)$ at timelike momentum 
transfers $s>4m_\pi^2$ is needed
to calculate the hadronic loop contribution 
to the muon anomalous magnetic moment and to the running of the e.m. coupling
(see \cite{g2,g21} for the current status). 

New accurate data on the pion form factor in the timelike region 
have recently been obtained by the 
CMD-2 collaboration \cite{CMD2002} measuring the 
$e^+e^-\to \pi^+\pi^-$ cross section at $\sqrt{s}= 0.61\div 0.96$ GeV
(for an update of these data see \cite{CM2update}). 
In this region, it is successfully described (fitted) 
using models based 
on $\rho$-meson dominance, with a small but clearly visible 
$\omega$-meson admixture. Above 1~GeV  
data on  $e^+e^-\to \pi^+\pi^-$ \cite{ADONE,Barkov,DM2} 
exist but are not that accurate.
Employing isospin symmetry  one also gains independent 
information from the measurements of
$\tau \to \pi^-\pi^0 \nu_\tau$ at $s<m_\tau^2$
(for details see, e.g. \cite{g2}). 
There are other interesting form factors 
closely related to $F_\pi$: the charged (neutral) kaon
e.m. form factors 
measured in $e^+e^-\to K^+K^-$($e^+e^-\to K^0\bar{K}^0$), 
as well as the weak transition form factor accessible in
$\tau \to K^- K^0 \nu_\tau$.

In the near future the experimental knowledge on $F_{\pi,K}(s)$ 
will be substantially improved, due to new data 
to be obtained using the radiative return method \cite{BinnKuhnMeln}.  
The first measurements of $F_\pi$ 
with this  technique in the $\rho$ region   
have already been performed by the KLOE Collaboration \cite{KLOE} 
and the agreement with the CMD-2 data is encouraging.
At larger energies, up to 2.0-2.5~GeV, perhaps even 3~GeV,
accurate measurements of $F_{\pi,K}$ are anticipated from the
BABAR experiment (for preliminary results see \cite{BABAR}). 
The high rates expected 
at $s\gg m_\rho^2$  demand phenomenological models 
more elaborated than the simple
$\rho$- ($\rho,\omega,\phi$-) dominance models for $F_\pi$ ($F_K$).
The main purpose of this paper is to construct an ansatz for
the pion form factor that is valid in the region below and far 
above $\rho$-resonance 
and to extend this model to the kaon form factor.

The model is constructed to obey the constraints from analyticity and
isospin-symmetry and to incorporate the proper behaviour at high
energies, consistent with perturbative QCD, and the correct
normalization at $s=0$. Furthermore it is based on plausible
assumptions derived from the quark model, moderate SU(3)-breaking, vector
dominance and a pattern of radial excitations expected from dual
resonance models. It has enough flexibility to accommodate the
characteristic interference pattern of the cross section and,
once sufficiently precise data are available at higher energies,
may be used to fix the parameters of the higher excitations. 

Above the $\rho$-resonance the excited $\rho'(1450)$ and $\rho''(1700)$ 
are expected to play an important role.
Already now these two states are indispensable, if one wants to
accommodate the measured parameters of the $\rho$-resonance with the
correct normalization of $F_\pi(s)$ at $s=0$, as demonstrated by 
earlier analyses and fits within the $\rho$ region 
(see, e.g. \cite{GT,KS}). In general,
to include all possible intermediate hadronic states in 
the $\gamma^*\to \pi^+\pi^-$ transition amplitude,
one has to take into account an infinite series of 
radially excited $\rho$'s. In addition, there are 
multihadron intermediate states with 
$J^P=1^-$ and $I=1$ ($2\pi, 4\pi, K\bar{K} $ etc.).  
Hence, the pion form factor at timelike $s>m_\rho^2$ is a 
complicated object determined by 
a large or even infinite amount of hadronic parameters 
not accessible at present in a rigorous theoretical framework.

On the other hand, at sufficiently large $s$ 
one expects $F_\pi(s)\sim \alpha_s(s)/s$, as predicted from 
perturbative QCD \cite{exclusive} for  
the pion form factor in the spacelike region at $s\to -\infty$,
analytically continued to $s \to +\infty$.
In other words, the overlap of many
intermediate hadronic states has to build up a smooth, 
power-behaved function. 
One might consider using this QCD prediction in the region of 
present interest, that is, at intermediate timelike $s$.  
However, data on the form factor in the spacelike
region, $s<0$, indicate that the onset of this asymptotic behaviour is  
far from the ``few GeV$^2$'' region. Preasymptotic contributions
$\sim 1/s^n$ with $n>1$, stemming from the end-point, soft mechanism 
\cite{endpoint} are essential for $s\sim 1-10 $ GeV$^2$.
Approximate methods valid at intermediate spacelike momenta, for example
QCD sum rules \cite{sr,lcsr,BK}, allow to calculate 
$F_\pi(s)$ including soft effects. However, a straightforward  
analytic continuation of $F_\pi(s<0)$ to large $s>0$ is difficult. 
The timelike form factor will suffer  
from uncertainties in  
the analytic continuation of soft parts, 
Sudakov logs and of $\alpha_s(s)$ (for a 
discussion see \cite{GP,BRS}). Hence,  QCD 
calculations cannot be directly used
in the large $s>0$ region, e.g., for 
estimating the ``tail'' of higher resonances 
in the form factor. 
In this paper we therefore prefer to adopt a model for the pion form-factor
formulated entirely in terms of hadronic degrees of freedom. 
Importantly, the form factor,  
as obtained from the model, fitted in the time-like region and
extrapolated to spacelike momenta, 
has a proper power-law behaviour which can be compared with the 
data at $s<0$ and used to test various QCD-based predictions. 

Models of hadronic amplitudes, where an infinite 
series of resonances at $s>0$ is summed to yield  
a power-law behaviour at $s<0$, are rooted in
the Veneziano amplitude and dual resonance models formulated long 
before the advent of QCD. 
Importantly, the pattern of infinite zero-width resonances 
is predicted in the $N_c=\infty$ limit of QCD.  
Recently, the model for the pion form factor 
using the masses and coefficients 
chosen according to Veneziano amplitude was considered in \cite{Dom}.
Earlier, similar analyses of the pion form factor can be found in
\cite{olddual}. Models of dual-resonance type with infinite 
number of resonances are widely used also for other hadronic problems, 
some recent works can be found in \cite{newdual}.
We will use the dual-QCD$_{N_c=\infty}$  model \cite{Dom} 
as a starting point,  modifying it 
for the first few $\rho$ resonances, by keeping their parameters 
(masses, widths and coefficients) free and fitting them 
to experiment. In this way, 
the complicated effects of $\rho$ resonances coupled   
to multihadron ($2\pi,4\pi$ etc.) states 
are implicitly taken into account.  
It is remarkable that the gross features of the model are well
reproduced with the (fitted) resonance parameters.
Since in the dual-resonance amplitude 
the coefficients of higher resonance contributions decrease
with the resonance number, 
the corresponding modifications for individual higher states
are not important, and the ``tail' of resonances 
is treated as in \cite{Dom}. 
The model for the pion form factor is also analytically continued to 
the spacelike region and compared with the data there and with the 
QCD predictions on $F_\pi(s)$ at large spacelike $s$.    

Furthermore, we extend the model to the kaon 
form factor, employing an SU(3)-generalization of the pion  
amplitude. Fitting the charged and neutral 
kaon form factors to the  data, and using flavour symmetries, 
we predict the weak kaon  
form factor relevant for $\tau$ semileptonic decays. 
Let us also mention at this point that the decomposition of the form factors
into their isospin zero and one components respectively, is also of relevance for
a model-independent evaluation of $\gamma$-Z-mixing \cite{Jegerlehner}
where the two amplitudes contribute with a different relative
weight.

The plan of the paper is as follows. In Section 2 we summarize the phenomenology
of $F_\pi(s)$  recalling the derivation of the  
$\rho$-meson contribution, whereas in Section 3 the contributions 
of excited $\rho$ resonances are discussed. The model for the pion
form factor is introduced in Section 4 and its parameters are fitted to the data.  
In Section 5 we proceed to the kaon e.m. form factors and 
Section 6 is devoted to the weak kaon form factor in $\tau$ decays.
Section 7 contains our summary and conclusions.

\section{The $\rho$-meson contribution to the pion form factor}

The pion e.m. form factor is defined in the standard way,
\be
\langle \pi^+(p_1)\pi^-(p_2)\mid j^{em}_\mu \mid 0 \rangle
=(p_1-p_2)_\mu F_\pi(s)
\label{formf}
\ee
The quark e.m. current
$
j^{em}_\mu=\sum\limits_{q=u,d,s} e_q \bar{q}\gamma_\mu q\,
\label{jem}
$
can be decomposed into isospin one and zero
components respectively. At the quark level this corresponds to
\be
j_\mu^{em} =\frac{1}{\sqrt{2}} j_\mu^{3} + \frac{1}{3\sqrt{2}} j_\mu^{I=0} - 
\frac{1}{3} j_\mu^{s}
\ee
with
\be
j_\mu^3 = (\bar u\gamma_\mu u - \bar d \gamma_\mu d )/\sqrt{2}\,,~~
j_\mu^{I=0} = (\bar u\gamma_\mu u + \bar d\gamma_\mu d )/\sqrt{2}\,,~~
j_\mu^{s} = \bar s \gamma_\mu s\,.
\label{components}
\ee
The isotriplet partners of the current $j_\mu^3$ form the 
charged weak current
\be
j_\mu^-= (j^1_\mu+ij^2_\mu)/\sqrt{2}=\bar{u}\gamma_\mu d .
\label{charged}
\ee
In the isospin symmetry limit, the $I=0$ and  $s$-quark components 
of the current do not contribute to $F_\pi$. In Eq.~(\ref{formf}), 
$s=(p_1+p_2)^2$ is the timelike momentum 
transfer squared, $s \geq 4m_\pi^2$. The form factor $F_\pi(s)$,
being analytically continued to the spacelike region $s< 0$, 
corresponds to the
hadronic matrix element $ \langle \pi^+(p_1)\mid j_\mu^{em} \mid \pi^+(-p_2)\rangle$ 
related to Eq.~(\ref{formf}) by crossing-symmetry.

There are very 
few model-independent relations determining or constraining 
the pion form factor. One of them is the normalization condition for the pion electric charge, 
\be
F_{\pi}(0)=1\,.
\label{norm}
\ee
An important role is played by the dispersion relation,
\be
F_{\pi}(s)=\frac1{\pi}\!\!\int\limits_{4m_\pi^2}^\infty
\!\!\!ds \frac{\mbox{Im}F_{\pi}(s')}{s'-s-i\epsilon}\,.
\label{disp}
\ee
The  asymptotic behaviour
expected from perturbative QCD  \cite{exclusive},
\be
\lim\limits_{s \to -\infty} 
F_{\pi}(s) \sim \frac{\alpha_s}{s}\,,
\ee
allows for an unsubtracted dispersion relation.
The application of Eq.~(\ref{disp})
is based on an independent equation for the imaginary part of the 
form factor derived from the unitarity condition,
\be
2\mbox{Im}F_{\pi}(s)(p_1-p_2)_\mu =
\sum\limits_h \int d\tau_h \langle \pi^+(p_1)\pi^-(p_2)\mid h\rangle \langle h
\mid j^{em}_\mu|0\rangle^*\,,  
\label{unitar}
\ee
where all possible hadronic states $h$ with 
$J^{PC}(I^G)=1^{--}(1^{+})$ are inserted. Each term in the sum in 
Eq.~(\ref{unitar}) includes the integration over the phase space and 
the summation over 
polarizations of the intermediate state $h$. Since isospin symmetry is not
exact, there could also be a small admixture
of the isospin-zero $J^{PC}=1^{--}$ states, e.g. the $\omega$ and 
its radial excitations.
 
Experimental data  reveal that at low $s\leq$ 1 GeV$^2$  
the most important contribution to Eq.~(\ref{unitar}) stems from 
$\rho$ meson. The  $\rho$-meson decay constant: 
\be
\langle \rho^0 \mid j_\mu^{em}\mid 0 \rangle = 
\frac{m_\rho f_\rho}{\sqrt{2}}\epsilon^{(\rho)*}_\mu \,,
\label{vect}
\ee
and the strong $\rho\pi\pi$ coupling:
\be
\langle \pi^+(p_1)\pi^-(p_2) |\rho^0\rangle= 
(p_2-p_1)^\alpha\epsilon^{(\rho)}_\alpha g_{\rho \pi\pi}\,,
\label{coupl} 
\ee
(where $\epsilon^{(\rho)}$ is 
the $\rho$-meson polarization four-vector) determine
the $\rho$-contribution to the 
imaginary part of the pion form factor in the narrow width approximation:
\be
\mbox{Im}F^{(\rho)}_{\pi}(s)= \frac{m_\rho f_\rho}{\sqrt{2}}g_{\rho\pi\pi}\pi\delta(s-m_\rho^2)\,.
\ee
Substituting this into the dispersion relation (\ref{disp}) gives:
\be
F^{(\rho)}_\pi(s)=
\frac{m_\rho f_\rho g_{\rho\pi\pi}}{\sqrt{2}(m_\rho^2-s-i\epsilon)}\,.
\label{Fpirho}
\ee
The excited $\rho', ...$ resonances have contributions of the same form,
with the decay constants $f_{\rho',..}$ and strong couplings 
$g_{\rho'\pi\pi},...$.

The pion form factor constructed by adding up  the zero-width
$\rho,\rho',...$ -resonances 
is an oversimplified ansatz which cannot be used at $s>0$ where  
the experimentally observable large widths of these resonances
are important. The widths are generated 
by the contributions of multi-hadron states to the 
imaginary part of $F_\pi$, starting from the lowest two-pion state.
The contribution of the latter to the unitarity relation 
\be
2\mbox{Im}F^{(2\pi)}_{\pi}(s)(p_1-p_2)_\mu=
\int d\tau_{2\pi} (p'_1-p'_2)_\mu A_{\pi\pi}(s)F_\pi^*(s)
\,,
\label{unitar2pi}
\ee  
involves the two-pion phase space :
$$
d\tau_{2\pi}=\frac{d^3p'_1}{(2\pi)^3 2E'_1}\frac{d^3p'_2}{(2\pi)^3 2E'_2}
(2\pi)^4\delta^4(p'_1+p'_2-p_1-p_2)
$$
where $(p'_1+p'_2)^2=s$ and $A_{\pi\pi}(s)$ is the amplitude of 
the strong pion-pion P-wave elastic scattering:
\be 
A_{\pi\pi}(s)=\langle \pi^+(p_1)\pi^-(p_2) | \pi^+(p'_1)\pi^-(p'_2)\rangle_{I=1,J=1}\,.
\label{2piampl}
\ee
In the low-energy region $4m_\pi^2<s< 16 m_\pi^2$, 
the two-pion state is the only contribution to the unitarity relation~\footnote{For a review on the low-energy pion form factor see \cite{Leutw}.}.
At larger $s$, vector mesons and various multihadron states contribute, 
making a model-independent use of Eq.~(\ref{unitar}) impossible.

A well-known and experimentally supported approach which we adopt here 
is ``vector dominance''. 
In its simplest form it assumes that the $\rho$ resonance saturates the
pion form factor (Eq.~(\ref{Fpirho})), thus requiring 
$f_\rho g_{\rho\pi\pi} /\sqrt{2}m_\rho =1$.
One furthermore approximates 
the pion-pion scattering amplitude with 
an intermediate $\rho$ exchange. Inserting the intermediate 
$\rho$ propagator in Eq.~(\ref{2piampl}) 
and using the definition (\ref{coupl}) we obtain:
\ba
A_{\pi\pi}(s)\simeq \frac{\langle \pi^+(p_1)\pi^-(p_2) |\rho^0\rangle \langle \rho^0|\pi^+(p'_1)\pi^-(p'_2)\rangle}{m_\rho^2-s} =-\frac{g_{\rho\pi\pi}^2(p_2-p_1)\cdot(p'_2-p'_1)}{(m_\rho^2-s)}\,.
\label{a2pi}
\ea
Substituting this ansatz together with Eq.~(\ref{Fpirho}) 
in Eq.~(\ref{unitar2pi}),
and integrating out the two-pion phase space with the help of 
\be
\int d\tau_{2\pi}(p'_2-p'_1)_\alpha (p'_1-p'_2)_\mu= \left( g_{\alpha\mu}-
\frac{(p_1+p_2)_\alpha(p_1+p_2)_\mu}{s}\right)\frac{[\,p(s)]^3}
{3\pi\sqrt{s}}\,,
\label{int2pi}
\ee
where $p(s) =\frac12 (s-4m_\pi^2)^{1/2}$ is the pion momentum in
c.m. frame of two pions,
we finally transform  Eq.~(\ref{unitar2pi}) to
\be
\mbox{Im}F^{(2\pi)}_{\pi}(s)= \frac{m_\rho f_\rho}{\sqrt{2}(m_\rho^2-s)}
\left\{\frac{g_{\rho\pi\pi}^2[\,p(s)]^3}{6\pi\sqrt{s}} \right\}  \frac{g_{\rho\pi\pi}}{m_\rho^2-s}\,.    
\label{ansatzIm}
\ee
This formula justifies using the more general expression
\be
F_{\pi}^{(2\pi)}(s)=\frac{m_\rho f_\rho}{\sqrt{2}(m_\rho^2-s)}
{\cal A}^{(2\pi)} (s) \frac{g_{\rho\pi\pi}}{m_\rho^2-s}\,,
\label{Fpi2pi}
\ee  
which can be interpreted as a two-pion loop insertion in the $\rho$
meson propagator. The amplitude ${\cal A}^{(2\pi)}(s)$ has a real and
imaginary part. A natural approximation for $\mbox{Im}{\cal A}^{(2\pi)}$ 
is the expression in curly brackets in Eq.~(\ref{ansatzIm}).  
At $s=m_\rho^2$ it is normalized to the $\rho\to 2\pi$ width
\be
\mbox{Im}{\cal A}^{(2\pi)}(m_\rho^2)=
m_\rho\Gamma(\rho\to 2\pi)= \frac{g_{\rho\pi\pi}^2}{6\pi
 m_\rho}\left [p(m_\rho^2) \right ]^3\,,
\label{gamrho}
\ee

To account for all possible amplitudes 
with two-pion insertions in the $\rho$-meson propagator,  
Eq.~(\ref{Fpi2pi}) has to be added 
to Eq.~(\ref{Fpirho}), together 
with subsequent  diagrams with two, three, etc. two-pion loops. 
Summing up this geometrical series yields:
\be
F_\pi^{(\rho+2\pi)}(s)=\left(
\frac{f_\rho g_{\rho\pi\pi}}{\sqrt{2}m_\rho}\right)
\frac{m_\rho^2}{m_\rho^2-s-\mbox{Re}{\cal A}^{(2\pi)}(s) 
-i\mbox{Im}{\cal A}^{(2\pi)}(s)}\,,
\label{BWgeneral}
\ee
for the part of $F_\pi$  which contains, 
in addition to the $\rho$-meson,
the contributions of the 2-pion intermediate state coupled
to $\rho$.

Several options for Eq.~(\ref{BWgeneral}) are in usage.
The simplest one is to neglect the real part of ${\cal A}^{(2\pi)}(s)$ and 
to  approximate the imaginary part by a constant, 
given by Eq.~(\ref{gamrho}). This gives the usual
Breit-Wigner (BW) formula for the $\rho$ resonance with a constant width. 
A more refined version (used e.g. in \cite{KS})  takes into account the 
$s$-dependence of Im${\cal A}^{(2\pi)}(s)$ in the form of the 
$p$-wave two-pion phase space (as indicated by Eq.~(\ref{ansatzIm}))
with the normalization from Eq.~(\ref{gamrho}):
\be 
\mbox{Re}{\cal A}^{(2\pi)}(s)=0,~~ 
\mbox{Im}{\cal A}^{(2\pi)}(s)= 
\sqrt{s}\frac{m_\rho^2}{s}\left(\frac{p(s)}{p(m_\rho^2)}\right)^3
\Gamma(\rho\to 2\pi)\equiv \sqrt{s}\Gamma_{\rho\to 2\pi}(s)\,.
\label{KSBW}
\ee
The function $\Gamma_{\rho\to 2\pi}(s)$ naturally
vanishes at $s<4m_\pi^2$, below the $2\pi$ threshold.

The Gounaris-Sakurai (GS) approach \cite{GS} represents another option 
widely used. In this case one takes into account  
a nonvanishing real part of ${\cal A}^{(2\pi)}(s)$ calculated  from 
the dispersion relation with two subtractions at $s=0$:
\be
{\cal A}^{(2\pi)}(s)= {\cal A}^{(2\pi)}(0)+ s \frac{d{\cal A}^{(2\pi)}(0)}{ds}+
\frac{s^2}{\pi}
\int\limits _{4m_\pi^2}^{\infty}\!ds' 
\frac{\mbox{Im}{\cal A}^{(2\pi)}(s')}{s'^2(s'-s-i\epsilon)}\,.
\label{doubledisp}
\ee
Using the expression for the imaginary part given in Eq.~(\ref{KSBW})
and changing the integration variable from $s^{(')}$ to
$v^{(')}=\sqrt{1-4m_\pi^2/s^{(')}}$ one transforms the integral in 
Eq.~(\ref{doubledisp}):
\ba
s^2\!\!\int\limits _{4m_\pi^2}^{\infty}\!ds' 
\frac{\mbox{Im}{\cal A}^{(2\pi)}(s')}{s'^2(s'-s-i\epsilon)}
=\left(\frac{m_\rho^2\Gamma(\rho\to 2\pi)}{8[p(m_\rho)]^3}\right)I(s)\,,
\nonumber
\\
 I(s)=
\frac{s}{v}\int_0^1 dv'v'^4\left[\frac{1}{v'-v-i\epsilon}-\frac{1}{v'+v}\right]\,.
\label{int}
\ea
Calculating the principal value of the integral 
yields the real part:
\be 
\mbox{Re} I(s)= s\left(\frac{2}{3}+2v^2 -v^3 \log\frac{1+v}{1-v}\right)\,.  
\label{int1}
\ee
Using the above result in Eq.~(\ref{doubledisp})
one obtains the real part of the amplitude ${\cal A}^{(2\pi)}$. 
Furthermore, following \cite{GS} 
the subtraction terms are fixed by 
the normalization conditions for the mass and the width of the $\rho$:
$Re{\cal A}^{(2\pi)}(m_\rho^2)=0$ and 
$\frac{d}{ds}Re {\cal A}^{(2\pi)}(m_\rho^2)=0$ . 
At $s=0$ the form factor is normalized to unity. Hence
\be
F_\pi^{(\rho+2\pi)}(s)=\left(
\frac{f_\rho g_{\rho\pi\pi}}{\sqrt{2}m_\rho}\right)
\frac{m_\rho^2+ H(0)}{m_\rho^2-s +H(s) 
-i\sqrt{s}\Gamma_{\rho\to 2\pi}(s)}\,,
\label{GS}
\ee
where we use the same notation as in \cite{KS}:
\be
H(s)= \hat{H}(s)-\hat{H}(m_\rho^2)-(s-m_\rho^2)\frac{d}{ds}\hat{H}(m_\rho^2)\,,\label{hh}
\ee
so that 
\be
\hat{H}(s)=\left(\frac{m_\rho^2\Gamma(\rho\to
    2\pi)}{2\pi[p(m_\rho)]^3}\right)
\left(\frac{s}4-m_\pi^2\right)v\log\frac{1+v}{1-v}\,.
\ee
From experiment $\Gamma(\rho\to 2 \pi)\simeq \Gamma_{tot}(\rho)$, hence
the couplings of $\rho$ to other intermediate states can be safely 
be neglected. Therefore we replace in both versions 
of the BW formula  the $\rho\to 2 \pi$ width 
by the total width removing the superscript $2\pi$ at the form factor. 

Finally, the $\rho$ contribution to the pion form factor 
introduced in Eq.~(\ref{Fpirho}) in the zero-width approximation 
and modified to include the width in Eq.~(\ref{BWgeneral}), 
can be rewritten in the following generic form:
\be
F_\pi^{(\rho)}(s)= c_\rho BW_\rho(s)
\label{Fpirho1}
\ee
where $c_\rho \equiv F_\pi^{(\rho)}(0)$ is the normalization
coefficient. In the adopted approximation is determined
by the product of $\rho$ decay constants and $\rho\pi\pi$ coupling:
\be
c_\rho=\frac{f_\rho g_{\rho\pi\pi}}{\sqrt{2}m_\rho}\,,
\ee
and $BW_\rho(s)$ is the BW formula 
normalized to unity at $s=0$. For this formula two different
versions will be used, one taken from \cite{KS}:
\be
BW_{\rho}^{KS}(s)=
\frac{m_\rho^2}{m_\rho^2-s 
-i\sqrt{s}\Gamma_{\rho}(s)}\,,
\label{KS1}
\ee
and the one from \cite{GS}
\be
BW_{\rho}^{GS}(s)=
\frac{m_\rho^2+ H(0)}{m_\rho^2-s +H(s) 
-i\sqrt{s}\Gamma_{\rho}(s)}\,.
\label{GS1}
\ee
In both cases the  effective $s$-dependent width is assumed to be
\be
\Gamma_\rho(s)=\Gamma_{\rho\to 2\pi}(s)\,,
\ee
with r.h.s. defined in Eq.~(\ref{KSBW}) and  $\Gamma_\rho(m_\rho^2)=
\Gamma^{tot}_\rho$.

\section{Contributions of excited $\rho$ states}

As already realized in earlier analyses of the pion form factor
(e.g., in \cite{GT,KS}), the single $\rho$-meson approximation 
is not sufficient to fulfil the normalization condition (\ref{norm}). 
Indeed, taking the measured values for $\Gamma(\rho\to l^+l^-)$ and 
$\Gamma(\rho\to 2 \pi)$  
from \cite{PDG} we obtain 
$f_\rho=220$ MeV  and, respectively  $g_{\rho\pi\pi}=6.0$,
yielding $c_\rho \simeq 1.2$. One needs 
to include the contributions of excited $\rho$ resonances to restore
the correct normalization. 
Currently, two of them, 
$\rho'(1450)$ and $\rho''(1700)$,
are well established experimentally \cite{PDG}.  
Adding the contributions of these two states in the form 
(\ref{Fpirho1}) to $F_\pi^{(\rho)}$,  
one fits experimental data on $e^+e^-\to 2 \pi$ in the 
$\rho$-region, practically up to
$s=1 $ GeV, restoring the normalization condition $F_\pi(0)=1$.
Both models (\ref{KS1}) and (\ref{GS1}) work well. 
In addition, there is a small 
isospin-violating 
effect from $\omega$ noticeable in the vicinity of $\rho$. In what follows, 
it will be taken into account as in \cite{KS}, 
by adding a $\rho-\omega$ mixing term to the $\rho$ contribution:
\be 
F_\pi^{(\rho)}(s) \to \frac{c_\rho BW_\rho(s)}{1+c_\omega}
(1+c_\omega BW_\omega)\,.
\label{omega}
\ee

Note, however, that the dominant decays of excited $\rho$'s are 
to final states other than $2\pi$,  hence the couplings of 
these resonances to various multiparticle intermediate states 
($4\pi$, $K\bar{K}$ etc.) become important at larger $s$, the region
of our interest.
In the previous section we have seen that the 
coupling of $\rho $ to the $2\pi$ state results in a  
geometrical series of $2\pi$-insertions  into the $\rho$ propagator
yielding an imaginary part normalized to the $\rho\to 2\pi$ width
in the formula for $BW_\rho$. The 
analogous summation  
procedure can be repeated for each multihadron state coupled 
to a given excited $\rho$, say, to $\rho'(1450)$. This leads to 
a formula for the $\rho'$ 
contribution to $F_\pi$ similar to  Eq.~(\ref{Fpirho1}),
where $\rho\to \rho'$ and 
the effective width in $BW_{\rho'}(s)$ 
is a sum over the effective widths for each channel
\be
\Gamma_{\rho'}(s)= 
\Gamma_{\rho'\to 2\pi}(s)+ \Gamma_{\rho'\to 4\pi}(s)+ 
\Gamma_{\rho'\to \rho 2\pi}(s)+ \Gamma_{\rho'\to K\bar{K}}(s)+... 
\label{gamma}
\ee
All we know about the functions on r.h.s. is their normalization
at $s=m_{\rho'}^2$ to the corresponding partial width of $\rho'$, 
so that altogether $\Gamma_{\rho'}(m_{\rho'}^2)=\Gamma^{tot}_{\rho'}$. 
The $s$-dependence 
for each partial width has to be introduced in a model-dependent
way, requiring detailed information  
on $4\pi$ (see e.g., \cite{CzK}) and other hadronic final states in $e^+e^-$.
In particular, to account for a proper threshold behaviour one has 
to introduce phase space factors for each $\Gamma_{\rho'\to f}$  
in Eq.~(\ref{gamma}), different from the $p$-wave factor for $\Gamma_\rho(s)$.
A complete kinematical and dynamical analysis of the partial widths 
for excited $\rho$ resonances is beyond our task (some models can 
be found in \cite{AchasovKozh}). 
We will continue using the same ansatz as for $\rho$,
with the simple $p$-wave threshold factor, having 
in mind that there is still a room for improvement  
at this point. We have checked that small modifications
of the effective width, e.g., replacing the effective 
threshold by $4m_\pi$ have little influence on the form factor.

Furthermore, the couplings of different vector resonances to 
one and the same multihadron state
generate mixing between these resonances. 
To give an example of this effect return to the unitarity relation 
(\ref{unitar2pi}) and substitute on r.h.s. the $\rho'$ resonance contribution 
to the form factor while keeping the intermediate $\rho$ exchange 
for $A_{\pi\pi}$. This term corresponds to a chain of transitions 
$\gamma^*\to \rho'\to 2\pi\to \rho\to 2\pi$. 
This nondiagonal amplitude is clearly not  
accounted for by the $2\pi$ insertions to the individual $\rho$- 
and $\rho'$-propagators. One has to add to the form factor 
new terms with the products of two BW propagators,
e.g. the $\rho'$ contribution to the form factor will have
the following schematic form:
\be   
F_\pi^{(\rho')}(s)= c_{\rho'}BW_{\rho'}(s)(1+x_{\rho'\rho}(s)BW_{\rho}(s)+...)
\label{mix}
\ee
where $x_{\rho'\rho}$ is the  mixing amplitude which is 
$s$-dependent, in general,
and ellipses indicate the mixing of $\rho'$ with other
$\rho$-resonances. 
Suppose one uses a specific dynamical model of $\rho$ resonances 
predicting the normalization factors $c_{\rho,\rho',..}$ of the 
BW-propagators.
After including the mixing, the amplitude
will have the form (\ref{mix}) with
a complicated $s$-dependence including an imaginary part. 
This system of mixed propagators could then be diagonalized, giving in
the general case rise to complex couplings between vector mesons and pions.
Since we do not attempt to solve this complicated dynamical pattern,
in the model of our choice we will keep 
the coefficients $c_{\rho,\rho',..}$  for few first  resonances      
as free fit parameters and for simplicity, restrict ourselves to real
values.

For the description of the infinite series of higher excitations we adopt an
ansatz rooted in the Veneziano amplitude and dual resonance models. The specific
{\em dual-QCD$_{N_c=\infty}$} amplitude, suggested in \cite{Dom}, contains
an infinite amount of zero-width vector mesons with 
the quantum numbers of $\rho$:
\be
F_\pi(s)=\sum\limits_{n=0}^{\infty}c_n\frac{m_n^2}{m_n^2-s}\,.
\label{Dom}
\ee
For convenience we will count 
$\rho$-resonances by a number $n$ which starts from $n=0$ 
for the $\rho$ meson, so that $\rho'(1450)$ and $\rho^{''}(1700)$
correspond to $n=1,2$, respectively.
The coefficients   
\be
c_n=\frac{(-1)^n \Gamma(\beta-1/2)}{\alpha'\sqrt{\pi}m_n^2\Gamma(n+1)
\Gamma(\beta-1-n)}
\label{cn}
\ee
decrease rapidly. The parameter $\alpha'=1/(2m_\rho^2)$ is related to the $\rho$-meson Regge
trajectory: $\alpha_{\rho}(s)=1+ \alpha'(s-m_\rho^2)$.
Furthermore, the model postulates an equidistant mass 
spectrum:
\be
m_n^2=m_\rho^2(1+2n)\,. 
\label{mn}
\ee
The parameter $\beta$ is free  and has to be fitted. 
We will use $c_0$  fitted from the $\rho$ region 
and calculate $\beta$ from Eq.~(\ref{cn}). 
As we shall see, the fit yields 
$c_0 = 1.098 - 1.171$ corresponding to 
$\beta= 2.16\div 2.3$, in agreement with \cite{Dom}. 

Using the above assumptions one easily obtains the form factor
in the closed analytical form:
\be
F_\pi(s)=\frac{\Gamma(\beta-1/2)}{\sqrt{\pi}\Gamma(\beta-1)}B(\beta-1,1/2-\alpha's)\,,
\label{Venez}
\ee
where $B(x,y)$ is Euler's Beta-function, so that $F_\pi(0)=1$. 
Importantly, in this model also the mean-squared charge radius 
of the pion (defined as 
$\langle r_\pi^2\rangle=6(dF_\pi(s)/ds)|_{s=0}$):
\be
\langle r_\pi^2 \rangle= 0.42\div 0.44~\mbox{fm}^2\,,  
\label{rpion}
\ee
at $\beta= 2.16\div 2.3$, agrees well with 
the experimental value \cite{Amendolia}  
$\langle r_\pi^2 \rangle_{exp}= 0.439\pm 0.008$ fm$^2$ and with the
recent chiral perturbation theory determination \cite{BijnensT} 
$\langle r_\pi^2 \rangle_{ChPT}= 0.452\pm 0.013$ fm$^2$.

The most spectacular property of the dual-QCD$_{N_c=\infty}$ model
(inherited from Veneziano amplitude) is
its explicit duality: in the timelike region the form factor 
has poles located at $s=m_n^2$,  whereas in the spacelike region 
at large $s<0$ it exhibits a smooth behaviour 
with a power-law asymptotic behaviour determined by the parameter $\beta$:
\be
\lim\limits_{s\to -\infty} F_\pi(s) \sim \frac{1}{s^{\beta-1}}\,.
\label{asympt}
\ee   
With $\beta=2.1-2.3$ this is very close to the prediction of 
perturbative QCD. 

In \cite{Dom} the model was further improved 
by including the constant widths of resonances
through the replacement:
\be
\frac{m_n^2}{m_n^2-s-i\epsilon} \to \frac{m_n^2}{m_n^2-s-im_n\Gamma_n}  \,.
\ee
The ansatz \cite{olddual,Dom} adopted for the total widths
is again motivated by string-like models:
\be 
\Gamma_n=\gamma m_n\,, 
\ee
with $\gamma=0.2$ fixed from $\rho$.
As explained above, to account for 
the presence of $2\pi$ and other intermediate multiparticle
states coupled to $\rho$ resonances 
one has to modify the widths to include $s$-dependence
with a proper threshold behaviour. 
Otherwise, the form factor predicts an unphysical 
imaginary part at $s=0$. Being unable to account for 
all possible intermediate multiparticle
states coupled to each $\rho_n$ 
we use, as a remedy, the threshold behaviour 
of $\rho_n\to 2 \pi$ partial width attributing it to the total width
\be
\Gamma_n(s)= \frac{m_n^2}{s}\left(\frac{p(s)}{p(m_n^2)}\right)^3\Gamma_n\,.  
\label{gamman}
\ee

\section{The model for the pion form factor} 

After all these modifications the  model \cite{Dom} 
for the pion form factor becomes
\be
F_\pi(s)= \sum\limits_{n=0}^{\infty}c_n BW_n(s)\,.
\ee
In its simplest form it depends only on few parameters 
($\beta$, $\alpha'$ and $\gamma$) and includes 
infinitely many hadronic degrees of freedom.

Importantly, the coefficients $c_n$  
decrease at $n\to \infty$. Hence
starting from $n\sim 4,5$, moderate deviations of 
$c_n,m_n,\Gamma_n$ from the model
predictions  do not influence the form factor, 
at least in the region of our interest, at $\sqrt{s}< 2-2.5$  GeV. 
On the other hand, for the most important first four resonances 
(including $\rho'''\equiv \rho_3$ with $m_3\simeq $2 GeV) we will allow  
the coefficients, masses and widths to deviate from the 
dual-QCD$_{N_c=\infty}$ model values, 
having in mind the effects of coupling to intermediate 
multiparticle states discussed in the previous section.  
To large extent, the fitted parameters will be close to those of the 
dual-QCD$_{N_c=\infty}$
model (see Table~\ref{tab:pion}).

Summarizing, our model for the pion form factor has the following form:
\be
F_\pi(s)= \left[\sum\limits_{n=0}^3 c_n BW_n(s)\right]_{fit}
+  \left[\sum\limits_{n=4}^{\infty}c_n BW_n(s)\right]_{dual-QCD_{N_c=\infty}}\,,
\label{themodel}
\ee
where in the $\rho$ contribution ($n=0$) the $\rho-\omega$ mixing is 
included
according to Eq.~(\ref{omega}) with the fixed parameters taken from \cite{KS}.
In the above, the parameters of the four lowest $\rho_{0,1,2,3}$ 
states (i.e. $\rho,\rho'(1450)$, $\rho''(1700)$ and $\rho'''$) are 
supposed to be fitted to experimental data, whereas the ``tail''
with the infinite amount of $\rho_{n>4}$ states   
is taken as in the dual-QCD$_{N_c=\infty}$ model.
However, having in mind the insufficient precision of the current data at $\sqrt{s}>$
1 GeV, we restrict the number of free fit parameters
to the coefficients $c_{0,1,2}$, 
the masses $m_0$ and $m_1$ and the total widths $\Gamma_0$ and $\Gamma_1$.    
The values of $\Gamma_2$ and $m_2$ are taken from \cite{PDG}. 
The coefficient $c_3$ of $\rho_3$ is fixed from 
the normalization condition  for the form factor: 
\be
c_3=1-(c_0+c_1+c_2)_{fit} -\left(\sum_{n=4}^{\infty} c_n
\right)_{dual-QCD_{N_c=\infty}}\,. 
\label{c3}
\ee
All remaining parameters in Eq.~(\ref{themodel}), 
that is $c_{n\geq 4}$, $m_{n\geq 3}$, 
and $\Gamma_{n\geq 3} $ are 
calculated from Eqs.~(\ref{cn}), (\ref{mn}) and (\ref{gamman}), respectively.  

Let us emphasize the main qualitative features of this model.
It nicely matches the existing $\rho$-dominance  
models at $\sqrt{s}<$ 1 GeV, such as the ones considered 
in \cite{KS}, simply because the $\rho_{n>2}$ states play a minor role
in the $\sqrt{s}<1$ GeV region. The model (\ref{themodel})
is flexible, that is, it allows to vary the proportion of fitted 
and modelled resonances above $\rho$.
E.g., with sufficiently precise data at higher energies one can include also the $\rho_4$-state  
into the 'fit' part. Alternatively, $\rho_{3}$ can be removed from the 'fit'
part and added to the dual-QCD$_{N_c=\infty}$ part. Furthermore, as 
mentioned already, $F_\pi(s)$, as given by Eq.~(\ref{themodel}), can be easily continued
to $s<0$ and compared with the experimental data
and QCD predictions in the spacelike region.
Accordingly, one gets a smooth power-like behaviour at 
asymptotically large timelike $s$. Altogether, the model contains a 
reasonably small amount of free parameters: three 
per each fitted resonance (the mass, coefficient and total width)
and three ``global'' parameters $\alpha',\beta,\gamma$,
 with $\alpha'=1/(2m_\rho^2$) taken from the Regge trajectory, 
$\beta$ taken from Eq.~(\ref{cn})
 with $c_0$ derived from the fit, and with $\gamma =0.2$. 
In principle, one can try  
to relax and independently fit also the three ``global'' parameters, 
but we prefer to keep them fixed.
Needless to say, the suggested ansatz has considerable room
for improvement, especially concerning the treatment
of the effective $s$-dependent resonance widths.   

We have fitted the model (\ref{themodel}) to the existing data 
\cite{CMD2002,ADONE,Barkov,DM2} 
for the timelike pion form factor. 
The results of the fit for the two different versions of 
BW-formulae: KS (Eq.~(\ref{KS1}) for all resonances) and GS (Eq.~(\ref{GS1})
for the first three resonances $\rho_{0,1,2}$), together with 
the relevant input parameters are presented in Table~\ref{tab:pion} and compared
with the predictions of the dual-QCD$_{N_c=\infty}$ model. 
\begin{table}[h]
\begin{tabular}{|c|c|r|r|r|r|}
\hline
Parameter & Input & Fit(KS)  & Fit(GS) & dual- & 
PDG value \\
&&&&QCD$_{N_c=\infty}$& \cite{PDG}\\
\hline
$m_\rho$          & - &  773.9\,$\pm$\,0.6& 776.3\,$\pm$\,0.6&input & 775.5\,$\pm$\,0.5 \\
$\Gamma_{\rho}$   & - & 144.9\,$\pm$\,1.0& 150.5\,$\pm$\,1.0 & input & 150.3\,$\pm$\,1.6 \\
\hline
$m_{\omega}$      & 783.0       & - & -    &  -  & 782.59\,$\pm$\,0.11 \\
$\Gamma_{\omega}$ & \,\,\,\,\,\,8.4 &-  &-    &  -  & 8.49\,$\pm$\,0.08 \\
\hline
$m_{\rho'}$        &-   & 1357\,$\pm$\,18 & 1380\,$\pm$\,18& 1335  &  1465\,$\pm$\,25 \\
$\Gamma_{\rho'}$   &- & 437\,$\pm$\,60& 340\,$\pm$\,53&  266    &  400\,$\pm$\,60 \\
\hline
$m_{\rho''}$       & 1700 &-  &- &1724 &  1720$\,\pm$\,20 \\
$\Gamma_{\rho''}$  & \,\,\,240  &-  &- & 344& 250\,$\pm$\,100 \\
\hline
$m_{\rho'''}$&  -   &-  &- & 2040 &-\\
$\Gamma_{\rho'''}$ &- &-  &- & 400& - \\
\hline
\hline
$c_0$        &-& 1.171$\pm$0.007&1.098$\pm$0.005&1.171 &-\\
$\beta$& $c_0$ and Eq.~(\ref{cn})
&2.30$\pm$0.01&2.16$\pm0.015$ & 2.3(input)&-\\
$c_\omega$&0.00184(KS)& -&-& -&-\\
&0.00195(GS)& & & &-\\
\hline
$c_1$&-&-0.119\,$\pm$\,0.011&-0.069\,$\pm$\,0.009&-0.1171&-\\
$c_2$&-&0.0115\,$\pm$\,0.0064&0.0216\,$\pm$\,0.0064&-0.0246&\\
$c_3$&Eq.~(\ref{c3})&-0.0438\,$\mp$\,0.02&-0.0309\,$\mp\,0.02$ &-0.00995&-\\
$\sum\limits_{n=4}^{\infty} c_n$& -0.01936
&-&-&-0.01936&-\\
\hline
$\chi^2/d.o.f.$&- &155/101& 153/101 &-&-\\
\hline
\end{tabular}
\caption{{\it Parameters of the pion form factor (\ref{themodel})
and results of the fit to the data. Masses and widths 
are given in MeV. The row 'Fit KS(GS)' contains the fitted values
for the case where all $BW_n$ are taken as in Eq.~(\ref{KS1})
($BW_{0,1,2}$ taken as in Eq.~(\ref{GS1})). The sum  $\sum\limits_{n=4}^{\infty} c_n$ is calculated from Eq.~(\ref{cn}). The PDG parameters 
for $\rho(770)$ are those listed in \cite{PDG} 
for ``Charged only, $\tau$ decays and $e^+e^-$''. The parameter $c_\omega$ is taken from \cite{KS}.}}
\label{tab:pion}
\end{table}
The results for $|F_\pi(s)|^2$ are plotted in Fig.~1, separately
for the $\sqrt{s}<1$ GeV  and $\sqrt{s}>1$ GeV regions.
\begin{figure}
\centering
\vspace{-1cm}
\includegraphics[width=0.7\textwidth]{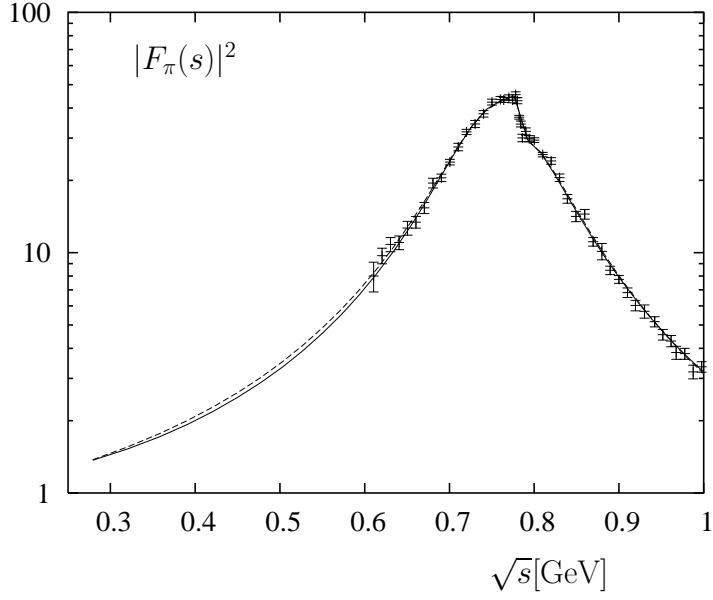}\label{fig-Fpitmla}
(a)
\\
\includegraphics[width=0.7\textwidth]{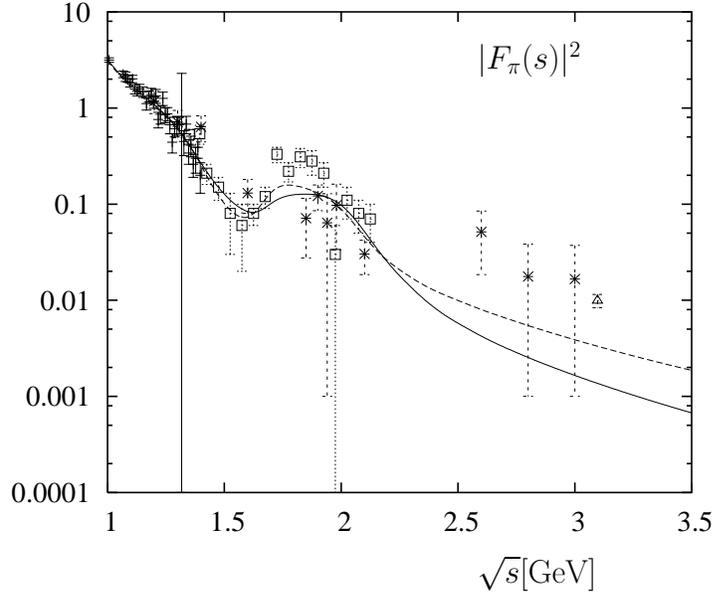}
\label{fig-Fpitmlb}
(b)
\caption{
\it The pion form factor squared $|F_\pi(s)|^2$ as a function of $\sqrt{s}$ 
fitted to the data in the region  near (a) and above (b) $\rho$ resonance.  
The solid (dashed) line
corresponds to the KS(GS) parameterization of BW formula. Data 
are taken from \cite{CMD2002,Barkov}(crosses),\cite{ADONE}(stars),
and \cite{DM2}(squares). The triangle point is the value of the form factor 
extracted from $J/\psi\to 2\pi$.}
\end{figure}

A few comments are in order.

The most spectacular result of the fit is the change of the 
$\rho''$ coefficient $c_2$  
with respect to the model \cite{Dom} and to the earlier
fits \cite{KS}, from negative values $\sim -(0.02-0.04)$ to smaller but 
positive values $\sim 0.01$-$0.02$. The positive sign is a direct consequence 
of the dip in the cross section around 1.6 GeV
(also the earlier fit of the data in \cite{DM2} revealed a similar pattern).

Furthermore, the fitted mass of $\rho'$ gets shifted  
with respect to the PDG value, the latter 
obtained by adding together data from all decay 
channels of $\rho'$. Note that a lower $m_{\rho'}$
consistent with our fit is also obtained by the fits in \cite{KS} 
and predicted by the dual-QCD$_{N_c=\infty}$ model. It is quite probable
that a more elaborated model of the total width of $\rho'$ 
including multiparticle thresholds will increase the fitted mass.  
We note that the values of the masses and widths of $\rho$, $\rho'$,
$\rho''$ as well as $c_0$
and $c_1$ are in the ballpark of the dual-QCD$_{N_c=\infty}$ model. 
The same is true for the magnitude of $c_2$, its positive sign is 
enforced by the dip around 1.6 GeV and might be a consequence of 
the strong mixing between $\rho''$ and nearby resonances.

In general, the $\rho'$, $\rho''$ and $\rho''$ terms 
and their interplay with the contribution of $\rho$
in both imaginary and real parts of the form factor 
are the main effects which determine the behavior
of $|F_\pi(s)|^2$ 
at $\sqrt{s}>1~\mbox{GeV}$. 
In particular, 
the dip in the form factor observed near $\sqrt{s}=1.6 $ GeV 
can only be described by  altering the sign of $c_2$.
The role of  the summed ``tail'' 
of $\rho_{n\geq 4}$ states is less important. 
One has to admit that the quality of the fit is not very high, 
(we get typically $\chi^2/d.o.f.\simeq 1.5$), in fact
this could simply indicate inconsistencies in the normalization of 
the various pieces of existing data in the timelike region. 
Moreover, the data points with large errors at $\sqrt{s}=2.5\div 3$ GeV 
are systematically higher than the fitted curve, and the value 
of $F_\pi(\sqrt{s}=m_{J/\psi})$ extracted from the $J/\psi\to 2 \pi$ 
partial width is larger than the model prediction 
by a factor of about three (this point was not included into the fit).
This disagreement deserves a comment.
Note that the extraction of $F_\pi(m_{J/\psi}^2)$ is 
``theoretically biased'', because one tacitly assumes that 
the intermediate photon exchange is the only mechanism in this
decay. Other mechanisms such as one-photon plus two
intermediate gluons could also be important in the $J/\psi \to 
2\pi$. Hence, the hadronic matrix element 
in this isospin-violating transition could be actually different 
from  the pion form factor. Although estimates \cite{Milana} of gluonic effects 
based on the perturbative charmonium annihilation 
are in favour of their smallness, we still think there is 
room for nonperturbative effects which are however not easily  
assessed. For the time being it seems difficult to accommodate a pion
form factor as large as the one derived from $J/\psi$ decay. 
(It will be interesting to check for a similar effect 
in the $4\pi$ mode.) 
\begin{figure}
\vspace{-1cm}
\centering
\includegraphics[width=0.7\textwidth]{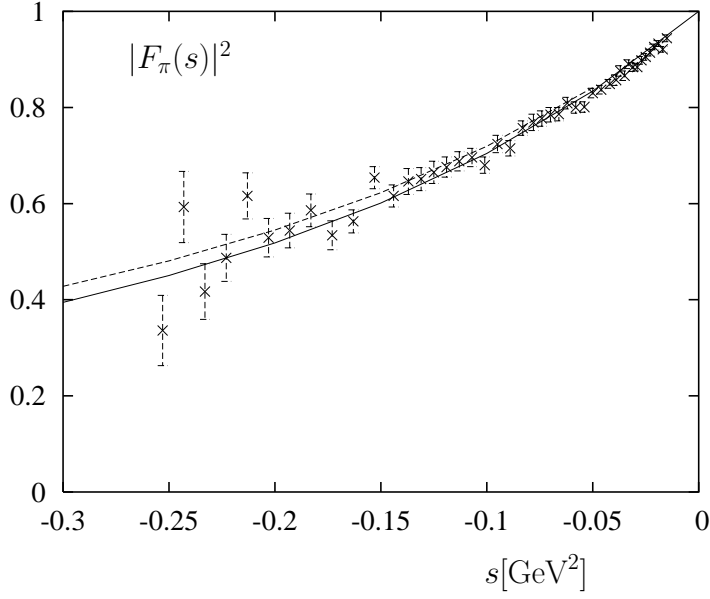}\label{fig-Fpispcla}
(a)
\\
\includegraphics[width=0.7\textwidth]{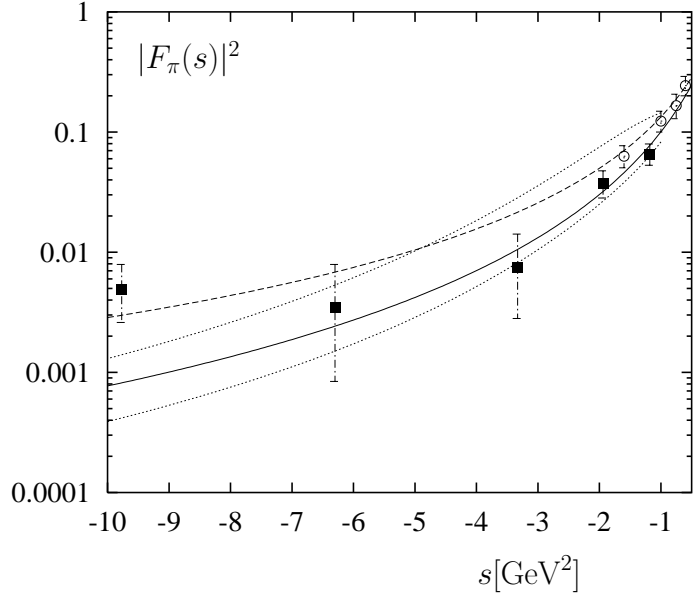}\label{fig-Fpispclb}
(b)
\caption{
\it The pion form factor squared 
$|F_\pi(s)|^2$ as a function of $s$ 
in the spacelike region at low $|s|$ (a) and large $|s|$ (b). 
The solid (dashed) line corresponds
to the analytic continuation of the timelike form factor 
(\ref{themodel}) with the KS(GS) parameterizations of the BW formula. Data 
are taken from \cite{Amendolia}(crosses), \cite{Jlab}(open circles) 
and \cite{Bebek} (full squares).
Dotted lines at $|s|> 1 \mbox{GeV}\,^2$ represent the interval derived from 
QCD light-cone sum rule predictions \cite{BK}.}
\end{figure}

To check our model further we continue $F_\pi(s)$  to $s<0$ and compare 
the result with the data there (see Fig. 2). The form factor 
obtained from direct electron-pion scattering  
at small $s<0$ (up to $|s|= 0.25~\mbox{GeV}^2$)~\cite{Amendolia} 
is not sensitive to the contributions 
of higher than $\rho$ resonances, provided the   
normalization to one at $s=0$ is imposed.
Note that the fitted form factor (\ref{themodel}) predicts
$\langle r_\pi \rangle^2 $= 0.440(0.426) fm$^2$ for the KS(GS) versions,
close to the dual-QCD$_{N_c=\infty}$ prediction (\ref{rpion})
and to the experimental value \cite{Amendolia} quoted above. 
  
At larger $|s|$ the old data \cite{Bebek} obtained 
from pion electroproduction have large errors and suffer from 
some intrinsic uncertainties \cite{Carlson}. More accurate data 
obtained  recently at JLab \cite{Jlab} 
at $|s|<1.6 ~\mbox{GeV}^2$ are in agreement with the model, but not sensitive 
to the details of the fit.  
Furthermore, there is reasonable agreement 
between the model and the QCD light-cone sum rule (LCSR) predictions
\cite{lcsr,BK}. The latter are taken from \cite{BK} with their 
estimated theoretical uncertainty. The pion form factor 
calculated from 3-point QCD sum rules 
\cite{sr} in the region of their validity $|s|\sim 1-4~ \mbox{GeV}^2$ 
is within the LCSR interval and therefore not shown separately.

If we try to artificially enhance the form factor
at large timelike region, e.g., by enhancing
the contribution of the ``tail'' trying to fit also 
the point at $\sqrt{s}=m_{J/\psi}$,  the form factor at spacelike $s<0$  
increases correspondingly, and the general agreement between data 
and QCD sum rule predictions is lost.
We conclude therefore, just as before, that it is implausible for 
the form factor obtained on the basis of dual
resonance approach 
to reach values   $|F_\pi(s)|^2 \geq 0.01$ at 
$\sqrt{s}= 2.5\div 3 $ GeV without conflicting with 
the spacelike data and especially with QCD predictions. 
This statement is independent of many details
involved in the timelike form factor model and in the QCD calculations.
It is therefore extremely interesting 
to obtain new accurate data at least up to  $\sqrt{s}=2.5$ GeV  
to check this conjecture.

\section{Charged and neutral kaon e.m. form factors}

We now adopt the analogous strategy to describe the kaon form
factors. Combining information on $K^+K^-$ and $K^0 \bar K^0$ production
with constraints from isospin symmetry, and using assumptions deduced
from the quark model and the OZI rule, it is possible to separate the
$I=1$ and $I=0$ amplitudes in the form factor, and even the $\omega$- and
$\phi$-components of the $I=0$ part. The $I=1$ part can then be used to
predict the rate for $\tau\to \nu K^- K^0$.

The electromagnetic form factors for charged and neutral kaons
defined similar to Eq.~(\ref{formf}):
\be
\langle K^+(p_1) K^-(p_2) | j^{em}_\mu | 0 \rangle = (p_1-p_2)_\mu \,F_{K^+}(s)
\label{formfKpl}
\ee
\be
\langle K^0(p_1)\bar K^0(p_2) | j^{em}_\mu | 0 \rangle = (p_1-p_2)_\mu\,F_{K^0}(s)\,,
\label{formfK0}
\ee
obey the constraints 
\be
F_{K^+}(0)=1,~~F_{K^0}(0)=0\,.
\label{Knorm}
\ee
They can be separated into their isospin one and zero parts,
\be
F_{K^+(K^0)}=F_{K^+(K^0)}^{(I=1)} + F_{K^+(K^0)}^{(I=0)}.
\ee
From isospin invariance one derives
\be
F_{K^+}^{(I=0)}=F_{K^0}^{(I=0)},~~F_{K^+}^{(I=1)}=-F_{K^0}^{(I=1)},
\ee
and the $I=1$ part can furthermore be used to predict the charged
current matrix element:
\be
\langle K^+(p_1) \bar{K}^0(p_2) | j^-_\mu | 0 \rangle =
(p_1-p_2)_\mu 2F_{K^+}^{(I=1)}(s)\,.
\label{KplK0}
\ee
A simultaneous fit to the two electromagnetic form factors leads,
therefore, to a direct prediction for the rate for $\tau\to \nu K^-
K^0$, without any further assumption.

In the context of vector dominance, combined with the quark model, the
kaon form factors
are saturated by $\rho$, $\omega$, $\phi$
and their radial excitations, 
\be
F_K(s)=\sum_{V=\rho,\omega,\phi,\rho',\omega',\phi',...}\frac{\kappa_Vf_V g_{V K\bar{K}}m_V}{m_V^2-s-im_V\Gamma_V}\,,
\label{FPV}
\ee
and it is the $\rho$-mediated $I=1$ part which
enters both the e.m. and the charged current matrix elements. 
We define the decay constants of the vector mesons via
\be
\langle V|j_\mu^{em}|0\rangle =\kappa_V 
m_V f_V \epsilon_{(V)}^{\mu*}\,,
\label{vect1}
\ee
where $\epsilon_{V}$ is the polarization vector of $V$,
and the coefficients 
$\kappa_\rho=1/\sqrt{2}$ (see Eq.~(\ref{vect})), 
$\kappa_\omega=1/(3\sqrt{2})$ and
$\kappa_\phi=-1/3$ reflect the valence quark content
of these mesons corresponding to  ``ideal'' mixing: 
\be
\rho^0=\frac{\bar{u}u-\bar{d}d }{\sqrt{2}},~~
\omega=\frac{\bar{u}u+\bar{d}d }{\sqrt{2}},~~
\phi= \bar{s}s\,.
\label{content}
\ee
The strong coupling is defined as in Eq.~(\ref{coupl}):
\be
\langle K(p_1)\bar{K}(p_2)|V\rangle= (p_2-p_1)^\nu\epsilon_{\nu(V)}g_{VK\bar{K}}\,.
\label{coupl1}
\ee
In the flavour SU(3)-symmetry limit one evidently 
has 
\be
F_{K^+}(s)=F_{\pi}(s),~~ F_{K^0}(s)=0\,.
\label{su3}
\ee
Since this symmetry 
is broken by the  strange-nonstrange quark-mass difference,
one has to expect quite noticeable deviations from Eq.~(\ref{su3}), 
in particular the $K^0$ form factor does not vanish
at nonzero $s$. Moreover, since $BR(\phi\to K^+K^-)\sim
BR(\phi\to K^0\bar{K}^0)$, the neutral and charged kaon
form factors have almost equal magnitudes near the $\phi$ resonance. 
In fact, the SU(3)-breaking also manifests itself 
in the valence quark content of vector mesons given in Eq.~(\ref{content})
and in the mass splitting between  $m_\rho\simeq m_\omega$ and  
$m_\phi$. 

\begin{figure}[t]
\begin{center}
\includegraphics[width=0.6\textwidth]{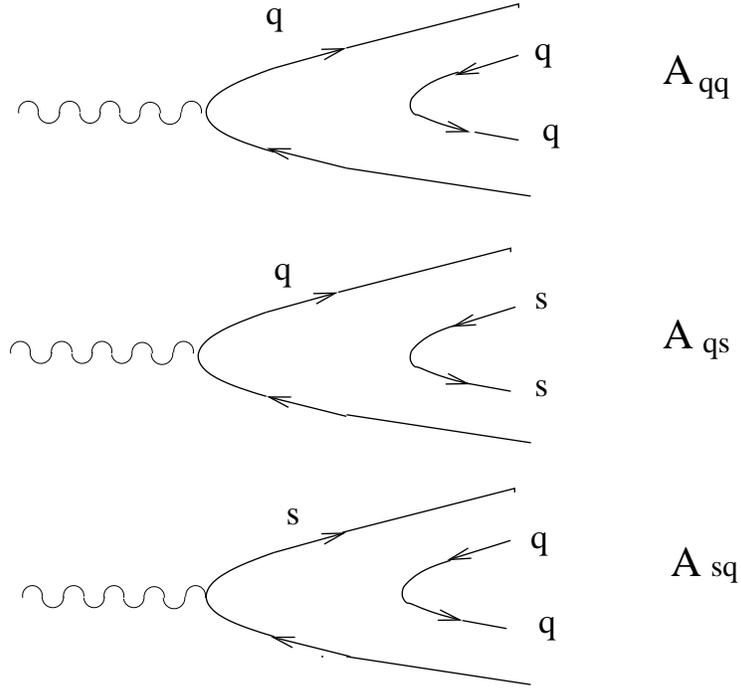}
\end{center}
\caption{{\em The quark diagrams corresponding to the 
contributions to the pion and kaon form factors with various flavour combinations. }} 
\end{figure}
In order to derive the kaon form factors in 
terms of separate vector meson contributions, it is 
convenient to consider a generic quark diagram of the 
strong $VP\bar{P}$ coupling and distinguish these diagrams
by the presence and position  of $s$ quarks. 
In the isospin symmetry limit, which we adopt here, there are 
three different diagrams depicted in Fig.~3:
1) without strange quarks (upper), 2) with $s$ and $\bar{s}$
in the $P\bar{P}$ state only (middle) and 3) with $s$ and $\bar{s}$
in the $V$ and the $P\bar{P}$ state (bottom). 
We denote the corresponding 
hadronic invariant amplitudes $A_{qq}$, $A_{qs}$ and $A_{sq}$. 
The diagrams with charge conjugated quark lines 
lead to the same amplitudes with an additional minus sign,
taking into account the negative C-parity of neutral vector mesons.
The strong couplings of $\rho,\omega,\phi$  
are then expressed in terms of diagrams:
\ba    
&&g_{\rho^0\pi^+\pi^-}\equiv g_{\rho\pi\pi}= \sqrt{2}A_{qq}\,,\nonumber\\
&&g_{\rho^0 K^+ K^-}= g_{\omega K^+K^-}=\frac{1}{\sqrt{2}}A_{qs}\,,~~
g_{\rho^0 K^0\bar{K}^0}=-g_{\omega K^0\bar{K}^0}= 
-\frac{1}{\sqrt{2}}A_{qs}\,,\nonumber\\
&&g_{\phi K^+K^-}=g_{\phi K^0\bar{K^0}}=- A_{sq}\,. 
\label{strongV}
\ea
Thus, from the
quark model one expects approximately equal $K\bar K \rho$ and $K\bar K
\omega$ couplings. It is easy to check that this simple formalism 
correctly reproduces $g_{\omega\pi^+\pi^-}=g_{\rho\pi^0\pi^0}=0$, 
as well as SU(3)-symmetry relations between the couplings.
In what follows we also use the relation between the 
decay constants of $\omega$ and $\rho$ following from 
Eq.~(\ref{vect1}): $f_\omega=f_\rho$. 
Substituting the decay constants and hadronic couplings
(\ref{strongV}) to Eq.~(\ref{FPV}) we obtain 
the desired decompositions of the kaon form factors 
in terms of vector meson contributions:
\be
F_{K^+}(s)= \frac{f_\rho A_{qs}}{2 m_\rho}BW_\rho(s)+
\frac{f_\rho A_{qs}}{6 m_\omega}BW_\omega(s)            
+\frac{f_\phi A_{sq}}{3 m_\phi}BW_\phi(s)\,,            
\label{Kformfpm}
\ee
\be
F_{K^0}(s)= -\frac{f_\rho A_{qs}}{2 m_\rho}BW_\rho(s)+
\frac{f_\rho A_{qs}}{6 m_\omega}BW_\omega(s)            
+\frac{f_\phi A_{sq}}{3 m_\phi}BW_\phi(s)\,.            
\label{Kformf0}
\ee
Written in the same terms the pion form factor is
\be
F_{\pi}(s)= \frac{f_{\rho} A_{qq}}{m_\rho}BW_\rho(s)\,,
\label{pionformf2}
\ee
and $f_\rho A_{qq}/m_\rho=1$ in the simplest version of VDM.
In the SU(3) limit
$f_\rho=f_\phi$, $m_\rho=m_\omega= m_\phi$ , $A_{qq}=A_{qs}=A_{sq}$
and the relations (\ref{su3}) are reproduced. In Eqs.~(\ref{Kformfpm}) and 
(\ref{Kformf0}) we will adopt the KS version of BW formulae for $\rho$, 
and the analogous $s$-dependent width for $\phi$, 
\be
\Gamma_\phi(s)= 
\frac{m_\phi^2}{s}\left(\frac{p^K(s)}{p^K(m_\phi^2)}\right)^3\Gamma_\phi\,.
\label{gammaphi}
\ee
with $\Gamma_\phi\equiv \Gamma^{tot}(\phi)$ and $p^K(s)=(s-4m_K^2)^{1/2}/2$.
For simplicity, we assume that the effective threshold of all 
$\phi$ decay modes including $\phi\to  3\pi$ is approximated
by Eq.~(\ref{gammaphi}), having in mind that 
the non-$K\bar{K}$ channels give only about 20\% of $\Gamma_\phi$. 
Naturally, the possibility to use the GS-form in 
Eqs.~(\ref{Kformfpm}) and (\ref{Kformf0}) exists, 
yielding inessential differences. For simplicity, to avoid complicated
$3\pi$-threshold factors we will use constant
widths for $\omega$, having in mind that the
thresholds are much lower than the boundary  of the physical region 
of the form factor: $9m_\pi^2\ll4m_K^2$. This is a good 
approximation at least for the narrow $\omega$ resonance.

Adding radial excitations to all ground-state
vector mesons is the next natural step. 
From the pion form factor analysis 
we already learned that the ``tails'' of higher resonances 
are numerically inessential. For the kaon form factor we therefore 
restrict the analysis to the 
excited states $\rho'$, $\rho''$,  
$\omega' \equiv \omega(1420)$ , $\omega''\equiv \omega(1650)$ 
and $\phi'\equiv \phi(1680)$ \cite{PDG}.
Higher excitations, as well as more elaborated $s$-dependent widths, 
can be installed in the future when more accurate data will be available. 
Since the products of decay constants and strong couplings 
in Eqs.~(\ref{Kformfpm}) and (\ref{Kformf0}) 
will not be separated  and have to be fitted as a whole,
it is convenient to introduce again the normalization factors
$c_V^K$ instead of these products. The ansatz 
for the kaon form factors
thus reads:
\ba
F_{K^+}(s)= \frac12 (c^K_\rho BW_{\rho}(s)+
c^K_{\rho'} BW_{\rho'}(s)+c^K_{\rho''} BW_{\rho''}(s))
\nonumber
\\+
\frac{1}{6}(c_\omega^K BW_\omega(s)+c_{\omega'}^K BW_{\omega'}(s)+
c_{\omega''}^K BW_{\omega''}(s))\nonumber \\
+\frac{1}3(c_{\phi} BW_{\phi}(s)+c_{\phi'} BW_{\phi'}(s))\,,
\label{Kformfpm1}
\ea
\ba
F_{K^0}(s)= -\frac12 (c^K_\rho BW_{\rho}(s)+
c^K_{\rho'} BW_{\rho'}(s)+c^K_{\rho''} BW_{\rho''}(s))\nonumber\\
+\frac{1}{6}(c_\omega^K BW_\omega(s)+c_{\omega'}^K BW_{\omega'}(s)+
c_{\omega''}^K BW_{\omega''}(s))
\nonumber \\
+\frac{1}3(\eta_\phi c_{\phi} BW_{\phi}(s)+c_{\phi'} BW_{\phi'}(s))\,,
\label{Kformf01}
\ea
The widths are with $p$-wave factors for $\rho$ and 
$\phi$ states as explained above, and constant for $\omega$-states,
which is however a rather crude approximation for $\omega',\omega''$. 
The ansatz in Eqs.~(\ref{Kformfpm1}) and (\ref{Kformf01}) reflects 
isospin invariance and the hierarchy 
of vector meson contributions according to their valence-quark
content, however, it allows for the possibility of SU(3) violations
which  could and will manifest in  differences between the fitted 
normalization coefficients.  
The additional factor $\eta_\phi$ in Eq.~(\ref{Kformf01})
takes into account the 
isospin-breaking difference between the charged and neutral kaon couplings
to $\phi$: 
\be
\eta_\phi\equiv \frac{g_{\phi K^0\bar{K}^0}}{g_{\phi K^+K^-}}=
\left(\frac{
BR(\phi\to K^0 \bar{K}^0)(m_\phi^2-4m_{K^+}^2)^{3/2}}{ 
BR(\phi\to K^+ K^-)(m_\phi^2-4m_{K^0}^2)^{3/2}} 
\right)^{1/2}\,.
\label{ratiophiKK}
\ee
According to \cite{PDG} the central value of this factor
slightly deviates from the unit:
\be
\eta_{\phi}= 1.027\pm 0.01\,.
\label{etaphiexp}
\ee
In the vicinity of the $\phi$ resonance this small effect
is noticeable in the fit, and as far as the branching ratio is concerned,
is dominated by the phase space factor.
The factor $\eta_\phi$ also takes care of Coulomb-rescattering 
and other isospin-violating differences
between charged and neutral modes (see also \cite{Voloshin,Bramonetal}).
To ensure the proper normalizations
$F_{K^0}(0)=0$ and $F_{K^+}(0)=1$, we introduce an additional
energy dependence with a simple step-function
\be  
\eta_\phi(s)= 1+(\eta_\phi-1)\theta(\sqrt{s}-(m_\phi-\Gamma_\phi))
\theta(m_\phi+\Gamma_\phi-\sqrt{s})\,,
\label{etaphi}
\ee
which in future, after this effect is better understood both experimentally
and theoretically, can be replaced by an appropriate analytical 
energy-dependence.

We have fitted the model (\ref{Kformfpm1})
and (\ref{Kformf01}) to the available data on charged 
\cite{AchasovK,AkhmetshinK,Ivanov,Dolinsky,BiselloK} and neutral 
\cite{CM2update,AchasovK,AkhmetshinK0,Mane} kaon form factors.
The masses and widths of $\rho$, $\omega$ and their excitations
are taken from \cite{PDG} and are listed in the Table~\ref{tab:kaon}.
Two different variants of the fit are carried out:

(1) the {\em constrained} fit 
(motivated by the quark model) where the normalization factors
for $\omega$ resonances are fixed: 
$c^K_{\omega,\omega',\omega''} = c^K_{\rho,\rho',\rho''}$ 
and only the normalization factors for the $\rho$ resonances are fitted;

(2) the {\em unconstrained} fit,  where $\omega$- and $\rho$- 
factors are fitted  as independent parameters.  

First, the mass and width of $\phi$ as well as the coefficient
$\eta_\phi$ are fitted in the region around $\phi$ resonance.
We obtain: $\eta_\phi= 1.011 \pm 0.009$ ($1.019 \pm 0.009$ ) for 
the constrained (unconstrained) fit, in a good agreement with the
experimental value (\ref{etaphiexp}).
Fixing $\eta_\phi$ and using $m_\phi$ and $\Gamma_\phi$ as starting values, 
the data  in the whole region of $\sqrt{s}$ are then fitted. 
The results of the fit are collected in Table~\ref{tab:kaon}.
\begin{table}
\begin{center}
\begin{tabular}{|c|r|r|r|r|}
\hline
Parameter & Input & Fit(1)& Fit(2) & PDG value\cite{PDG}\\
&&&&\\
\hline
$m_\phi$  & - & 1019.372\,$\pm$\,0.02&1019.355\,$\pm$\,0.02& 1019.456\,$\pm$\,0.02\,\\
$\Gamma_{\phi}$ & - &  4.36\,$\pm$\,0.05&4.29\,$\pm$\,0.05 & 4.26\,$\pm$\,0.05\\ 
\hline
$m_{\phi'}$          & 1680 &-&- &1680\,$\pm$\,20\\
$\Gamma_{\phi'}$ & 150 &- &-& 150\,$\pm$\,50\\ 
\hline
\hline
$m_\rho$        & 775 &- &-&775.8\,$\pm$\,0.5 \\
$\Gamma_{\rho}$ & 150 &- &- &150.3\,$\pm$\,1.6 \\
\hline
$m_{\rho'}$     & 1465 &-&- & 1465\,$\pm$\,25 \\
$\Gamma_{\rho'}$ & 400  &- &-&  400\,$\pm$\,60 \\
\hline
$m_{\rho''}$       & 1720 &- &-& 1720$\,\pm$\,20 \\
$\Gamma_{\rho''}$  & \,\,\,250 &-&- & 250\,$\pm$\,100 \\
\hline
$m_{\omega}$     & 783.0 &  -  &-& 782.59\,$\pm$\,0.11 \\
$\Gamma_{\omega}$ & \,\,\,\,\,\,8.4 &  -  &-& 8.49\,$\pm$\,0.08 \\
\hline
$m_{\omega'}$   & 1425 &  -  &-& 1400-1450  \\
$\Gamma_{\omega'}$ &215  &  -  &-& 180-250\\
\hline
$m_{\omega''}$ & 1670   &  -  &-& 1670\,$\pm$ 30\, \\
$\Gamma_{\omega''}$ & \,\,\,315 &-&  -  & 315\,$\pm$\,35 \\
\hline
\hline
$c_{\phi}$& -             & 1.018\,$\pm$\,0.006  &0.999\,$\pm$\,0.007  
&  -\\
$c_{\phi'}$& $1-c_\phi^K$ & -0.018\,$\mp$\,\,0.006 &0.001\,$\mp$\,0.007&-\\
\hline
$c_\rho^K$ &- &1.195\,$\pm$\,0.009 &1.139\,$\pm$\,0.010  &-\\
$c_{\rho'}^K$  & -& -0.112\,$\pm$\,0.010 & -0.124\,$\pm$\,0.012  &-\\
$ c_{\rho''}^K$& $1-c_\rho^K-c_{\rho'}^K$&-0.083\,$\mp$\,0.019 &-0.015\,$\mp$\,0.022   &-\\ 
\hline
\hspace{0.5cm}$c_\omega^K$(1)& $c^K_\rho$ &1.195\,$\pm$\,0.009 &-&-\\
\hspace{0.5cm}$c_{\omega}^K$(2)&-&-& 1.467\,$\pm$\,0.035&-\\
\hspace{0.5cm}$c_{\omega'}^K$(1)& $c^K_{\rho'}$ &-0.112\,$\pm$\,0.010&-&-\\ 
\hspace{0.5cm}$c_{\omega'}^K$(2)&-&-&-0.018\,$\pm$\,0.024 &-\\
$ c_{\omega''}^K$ & $1-c_{\omega }^K-c_{\omega'}^K$&-0.083\,$\mp$\,0.019 &-0.449\,$\mp$\,0.059 &-\\ 
\hline
$\chi^2/d.o.f.$&- & 328/242 &281/240&-\\
\hline
\end{tabular}
\caption{{\it Parameters of the kaon form factors 
and results of the fit to the data. Masses and widths 
are given in MeV. The row 'Fit(1)' (Fit(2))contains the values
of the constrained (unconstrained) fits.}}.
\label{tab:kaon}
\end{center}
\end{table}
The best (i.e., stable and physically plausible) results for both variants of the fit are obtained if  
data on $F_{K^+}$ and $F_{K^0}$ are fitted simultaneously.
Thus, predicting $F_{K^0}$ from $F_{K^+}$ with the 
currently available data is not yet possible. 
The resulting curves for the form factors are plotted in Figs.~4,5. 
Most importantly, fitting the kaon form factor
above $\phi$ resonance, it is  indeed possible
to extract separate $\rho,\omega,\phi$ components,
which was not possible in the $\phi$ region due to 
the dominance of this resonance.     

We also  find the pattern of the normalization factors
$c_{\rho,\rho'}^K$  for the first two $\rho$-resonance 
to be very similar
to the corresponding values $c_{0,1}$ obtained in the pion form factor fits. 
These factors can be immediately translated into the strong 
couplings dividing out the decay constants of vector mesons.
The latter are independently measured in the leptonic decays revealing 
a very mild SU(3) breaking, at the level of 10 \%;
according to the data in \cite{PDG}: $f_\rho\simeq$ 220  MeV,  
$f_\omega\simeq$ 195  MeV and   $f_\phi\simeq$ 228  MeV.  
The SU(3)-violating difference between the couplings 
of $\rho$ and $\phi$ to kaons
estimated from comparing $c_\rho^K$ and $c_\phi$  
is also moderate in both versions of the fit. 

As already noticed above, the constraint $c_\omega^K=c_\rho^K$ 
naturally follows from the valence quark content of both mesons 
and we consider this constraint as a part of our model.
The fact that the unconstrained fit gives about 25\%  difference
between these two coefficients,  
a noticeable deviation from the quark-diagram relation,
should be taken with caution, having in mind poor quality of data.
Also $\chi^2$'s of both fits are in the same ballpark,
so that from the fitting point of view we cannot yet give 
any preference to the version with the 
``floating'' couplings of $\omega$-resonances.
On the other hand, this difference indicates  that the fit is able 
to resolve also the ``fine structure'' of the couplings.  
The differences between the normalization factors 
given by the fit for excited resonances: $c_{\rho'}^K$ 
vs. $c_{\omega'}^K$ (in the unconstrained
fit), $c_{\rho'}^K$ vs. $c_{\phi'}$, etc. 
are generally large, which is not surprising,
in view of the complicated mixing between all these states.
Including in the future more precise data and switching on the ``tails'' of
the dual QCD$_{N_c=\infty}$ amplitudes in the kaon form factors
for all three vector mesons  will allow to reveal
these differences more accurately. 

Furthermore, an indication for an excess of the measured charged kaon 
form factor vs. the model is present in Fig.~4b in the region around 2 GeV,
although the experimental errors are large. Remember, that we have not
included in our fit the contribution of the second excited 
$\phi''$ state with a mass around 2 GeV, which might be responsible for this 
potential difference. However, we refrain from further investigation 
before  more accurate data are available.

\begin{figure}
\vspace{-1cm}
\centering
\hspace{-0.6cm}
\includegraphics[width=0.7\textwidth]{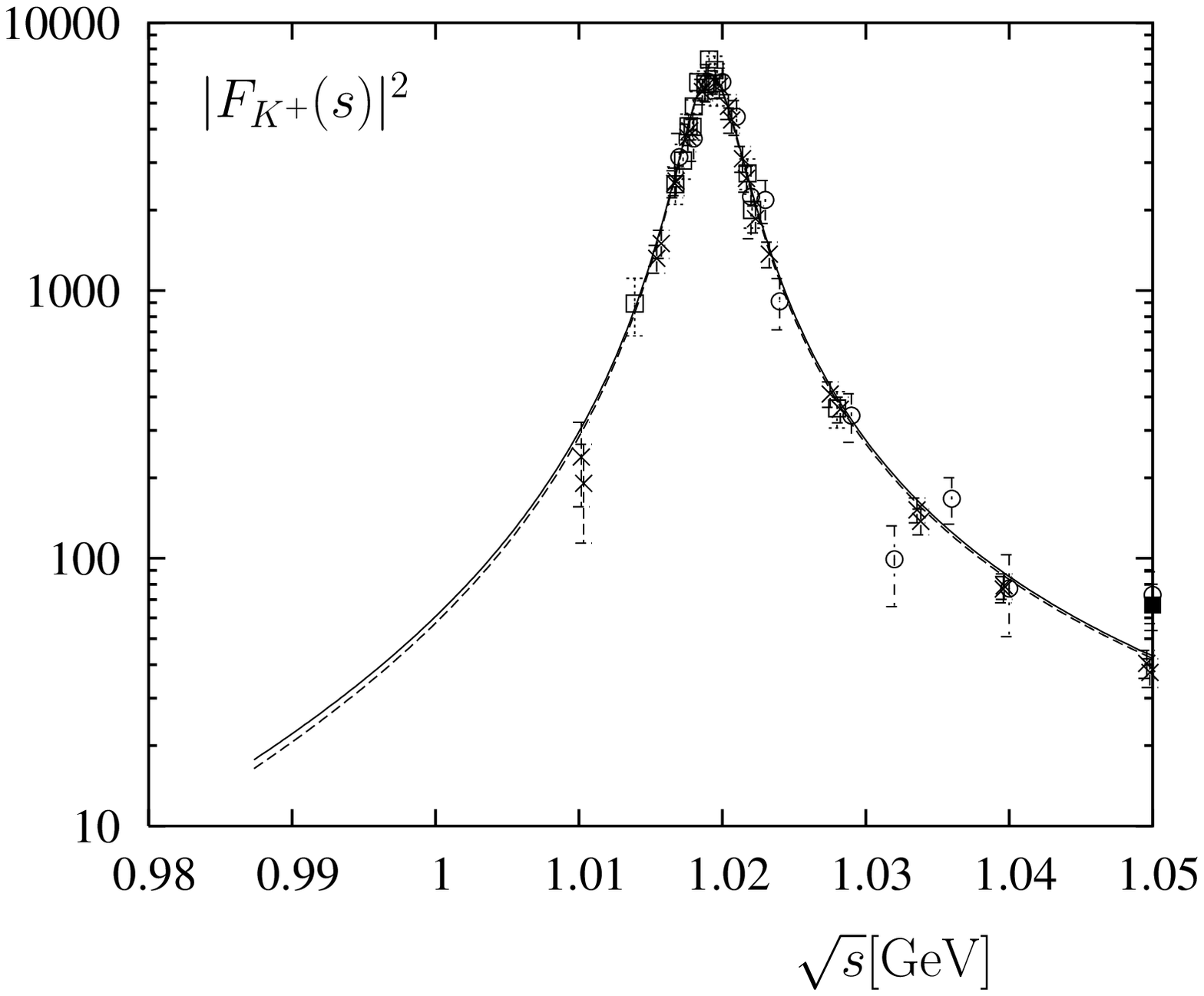}\label{fig-FKpla}
(a)
\\
\includegraphics[width=0.7\textwidth]{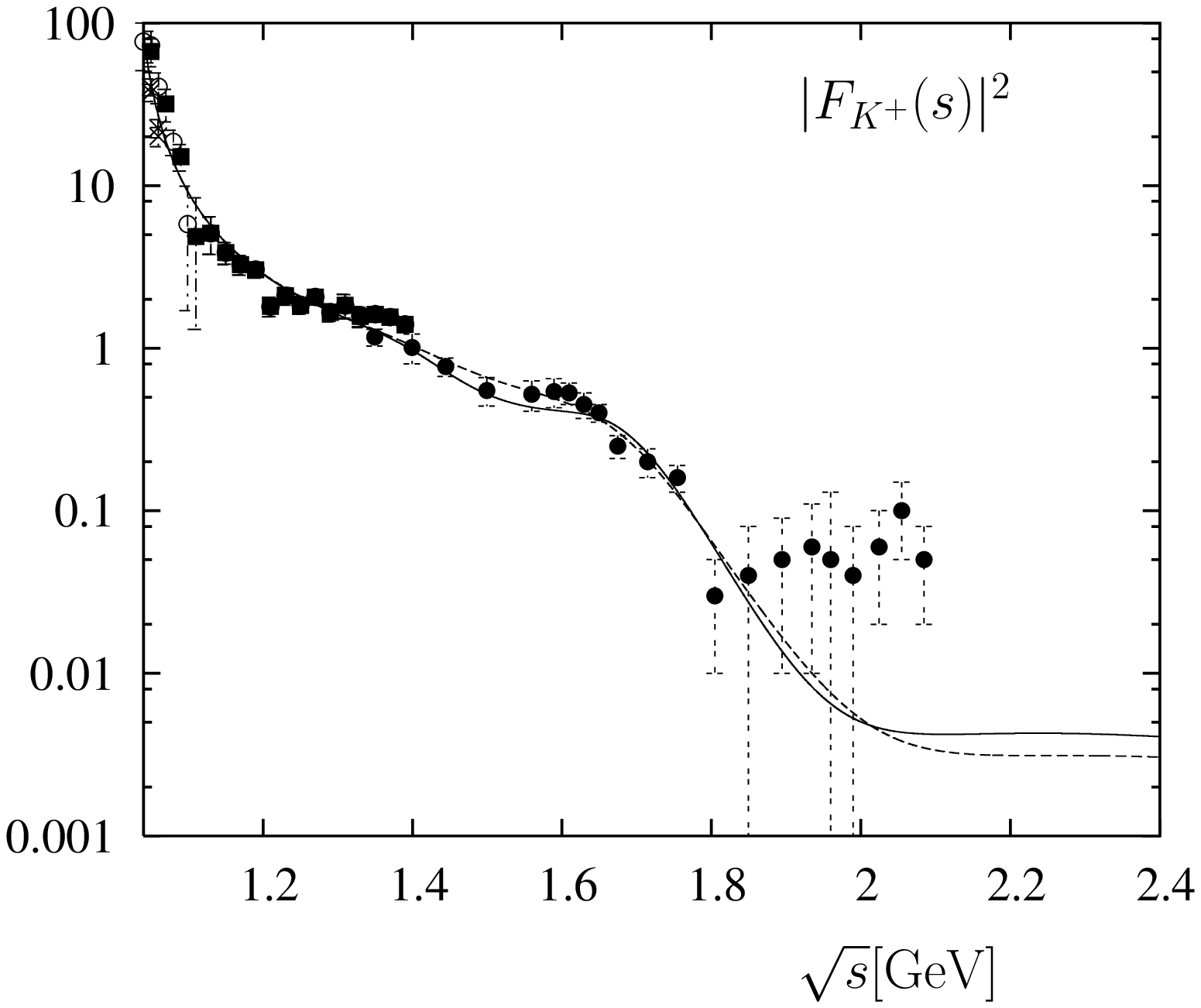}\label{fig-FKplb}
(b)
\caption{
\it 
The charged kaon form factor squared 
$|F_{K^+}(s)|^2$ as a function of $\sqrt{s}$ 
fitted to the data taken from \cite{AchasovK} (crosses),
\cite{AkhmetshinK} (open squares), \cite{Ivanov} (open circles),
\cite{Dolinsky} (full squares) and \cite{BiselloK} (full circles). 
The solid(dashed) lines correspond to the constrained (unconstrained)
fit.}
\end{figure}

\begin{figure}
\vspace{-1cm}
\centering
\includegraphics[width=0.7\textwidth]{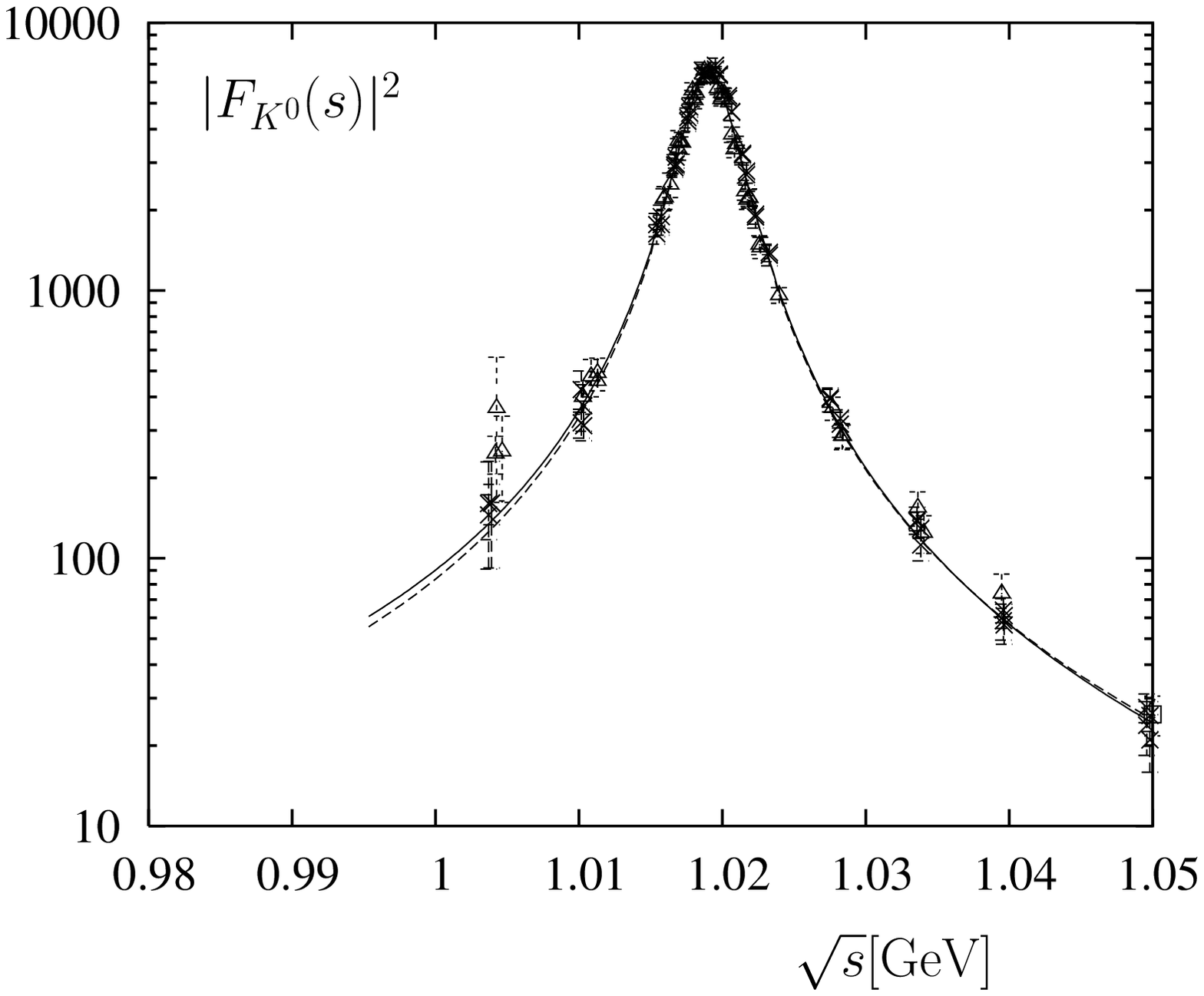}\label{fig-FK0a}
(a)
\\
\includegraphics[width=0.7\textwidth]{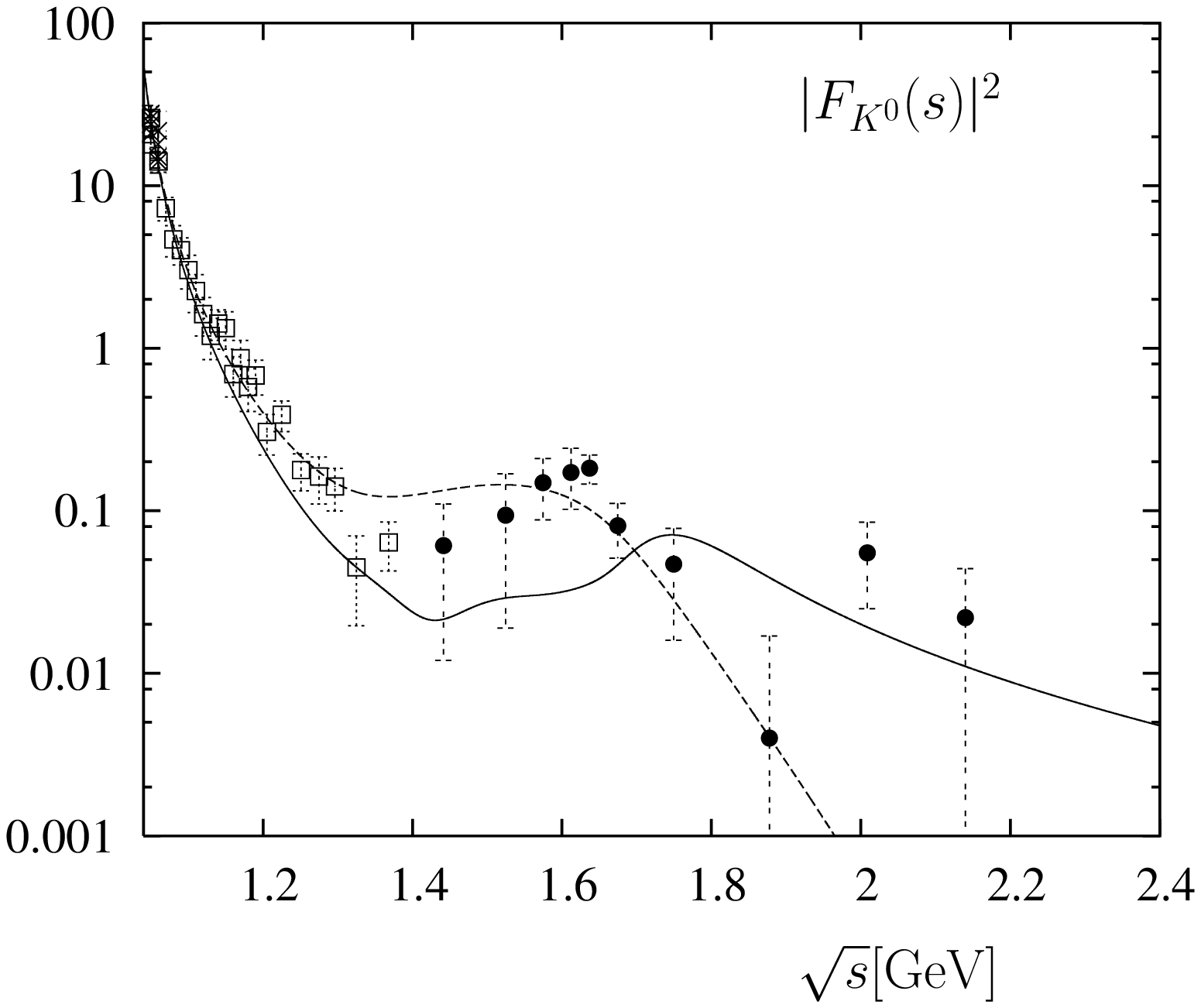}\label{fig-FK0b}
(b)
\caption{
\it 
The neutral kaon form factor squared 
$|F_{K^0}(s)|^2$ as a function of $\sqrt{s}$ 
fitted to the data taken from \cite{CM2update}(triangles), \cite{AchasovK}(crosses),
\cite{AkhmetshinK0}(open squares) and 
\cite{Mane}(full circles). The solid(dashed) lines correspond to the constrained (unconstrained)
fit.}
\end{figure}
The mean-squared charge radius of the $K^+$ obtained 
in our model: $\sqrt{\langle r_K^2\rangle}= 0.56$ fm
(for both fits and with a small error), is in a good
agreement with the experimental value \cite{Dally} 
 $\sqrt{\langle r_K^2\rangle}_{exp}= 0.53\pm 0.05$ fm.
We have also checked that, being analytically continued to 
large $s<-1 \mbox{GeV}^2$, the charged kaon form factor agrees
with the LCSR prediction obtained in \cite{BK}.   

Finally, as mentioned in the Introduction, the separate 
reconstruction of $I=0$ and $I=1$ spectral functions
might be a useful ingredient for various phenomenological
analyses. In Fig.~6
we display  
the spectral functions defined as
\be
\rho_{K\bar{K}}^{(I=0,1)}(s)=
\frac{1}{12\pi}
\left| \frac{F_{K^+}(s)\pm F_{K^0}(s)}{2}\right|^2
\left(\frac{2p_K(s)}{\sqrt{s}}\right)^3\,,
\label{spectral}
\ee
noticing that this observable is quite sensitive to the 
pattern of resonances in the form factor.
\begin{figure}
\vspace{-5cm}
\centering
\hspace{-0.8cm}
\includegraphics[width=0.75\textwidth]{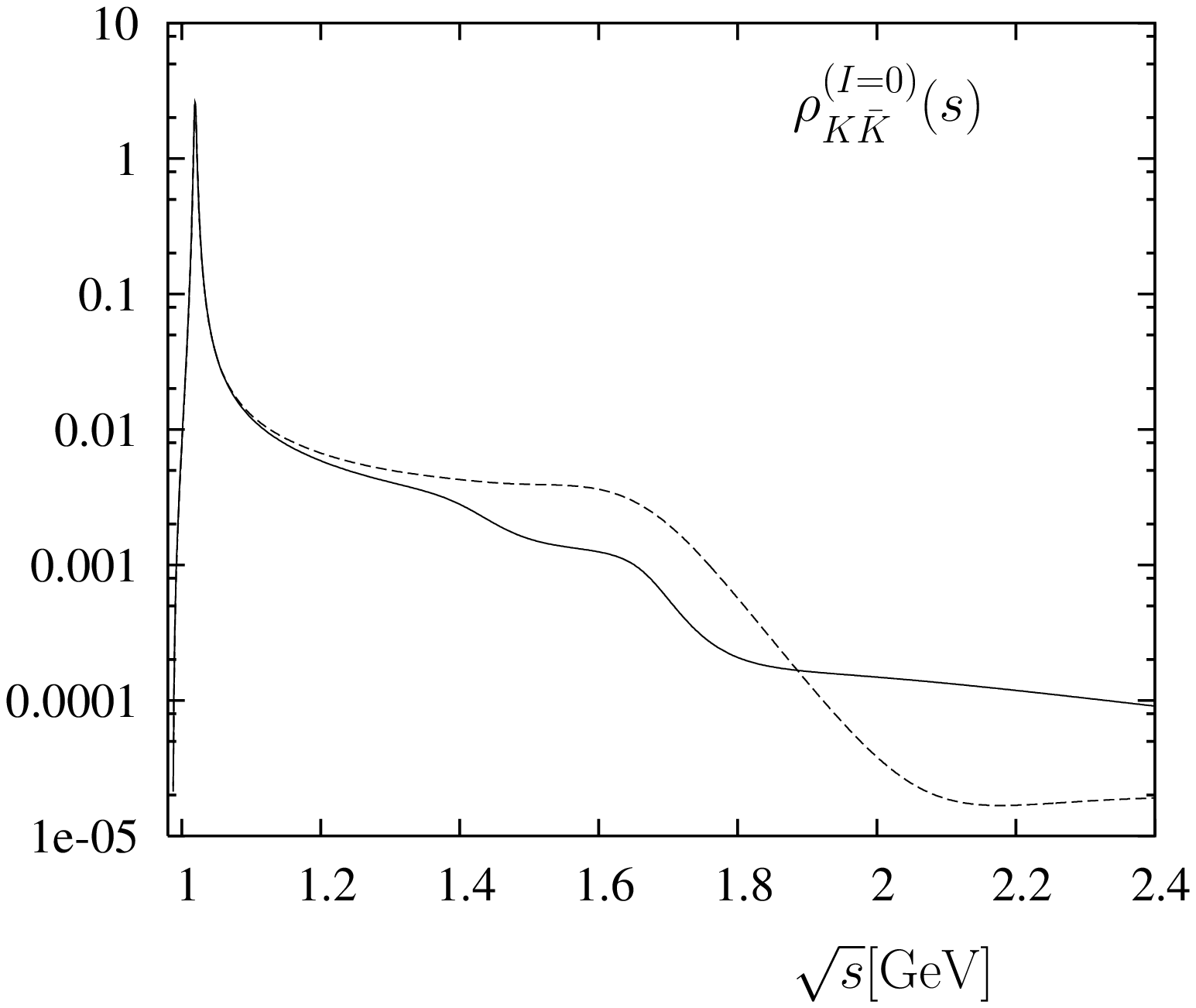}\label{fig-spectr_a}
(a)
\\
\includegraphics[width=0.7\textwidth]{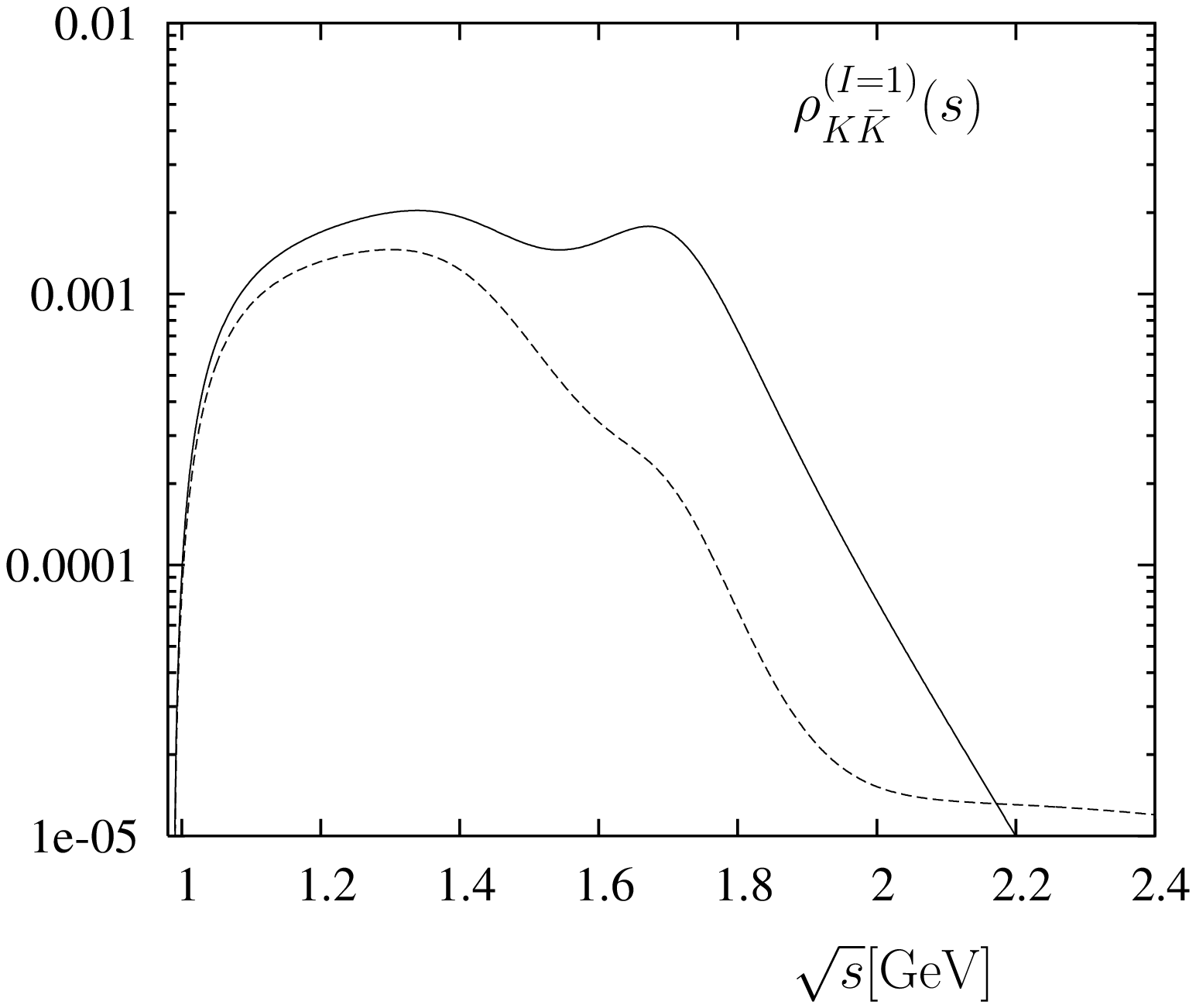}\label{fig-spectr_b}
(b)
\caption{
\it 
The spectral functions (\ref{spectral}) with $I=0$ (a) 
and $I=1$ (b) obtained from the fitted kaon form factors.
The solid(dashed) lines correspond to the constrained (unconstrained)
fit.}
\end{figure}


\section{Predicting $\tau\to K^- K^0\nu_\tau$ decay distribution and rate}

As emphasized above, the isospin one part of the e.m. kaon form factor, together
with the isospin-symmetry relation
(\ref{KplK0}), can be used to predict the $\tau\to K^- K^0\nu_\tau$ decay width.
The differential decay distribution in $\sqrt{Q^2}$ 
(the invariant mass of the kaon pair), 
normalized to the leptonic width of $\tau$  reads:
\ba
&&\left(\frac{1}{BR(\tau\to \mu^-\bar{\nu}_\mu \nu_\tau)}\right)
\frac{dBR(\tau\to K^-K^0\nu_\tau)}{d\sqrt{Q^2}}
\nonumber\\
&&=
\frac{|V_{ud}|^2}{2 m_\tau^2}
\left(1+\frac{2Q^2}{m_\tau^2}\right)\left(1-\frac{Q^2}{m_\tau^2}\right)^2
\left(1-\frac{4m_K^2}{Q^2}\right)^{3/2}\sqrt{Q^2}\,|F_{K^-K^0}(Q^2)|^2\,.
\label{distr}
\ea
In accordance with the isospin limit, we neglect the mass
difference between charged and neutral kaons and the 
effect of the scalar form factor.
Using Eq.~(\ref{KplK0}) we have $ F_{K^-K^0}=-2F_{K^+}^{(I=1)}$, 
hence 
\be
|F_{K^-K^0}(Q^2)|^2= |c^K_\rho BW_{\rho}(Q^2)+
c^K_{\rho'} BW_{\rho'}(Q^2)+c^K_{\rho''} BW_{\rho''}(Q^2)|^2\,.
\label{FKminK0} 
\ee
The fitted values for $c_{\rho,\rho',\rho''}^K$
from Table~\ref{tab:kaon} thus allow us to 
calculate the decay distribution (\ref{distr}).
The normalized distribution is plotted in Fig.~7 and 
(qualitatively) compared with the 
event distribution in the kaon pair mass,
measured by CLEO Collaboration \cite{CLEOtau}. Another measurement
of this distribution by  ALEPH Collaboration can be found in \cite{ALEPHtau}.
Integrating over $\sqrt{Q^2}$ from $2m_K$ to $m_\tau$ 
we obtain the branching ratio
\be
BR(\tau\to K^-K^0 \nu_\tau )= 0.19\pm 0.01\% ~~(0.13 \pm 0.01 \%)  
\ee
for the constrained (unconstrained) fit,
to be compared  with the experimentally measured value \cite{PDG}
\be
BR(\tau\to K^-K^0 \nu_\tau)= 0.154\pm 0.016\%.  
\ee
We see that both the decay distribution and the decay width
are very sensitive to the pattern of $\rho$ resonances
in the isospin-1 form factor. Generally, the width grows
with the increase of the excited $\rho$ contributions, 
an effect observed earlier in \cite{FKM} (see also \cite{EidI}).   

\begin{figure}
\vspace{-1cm}
\centering
\hspace{-0.6cm}
\includegraphics[width=0.7\textwidth]{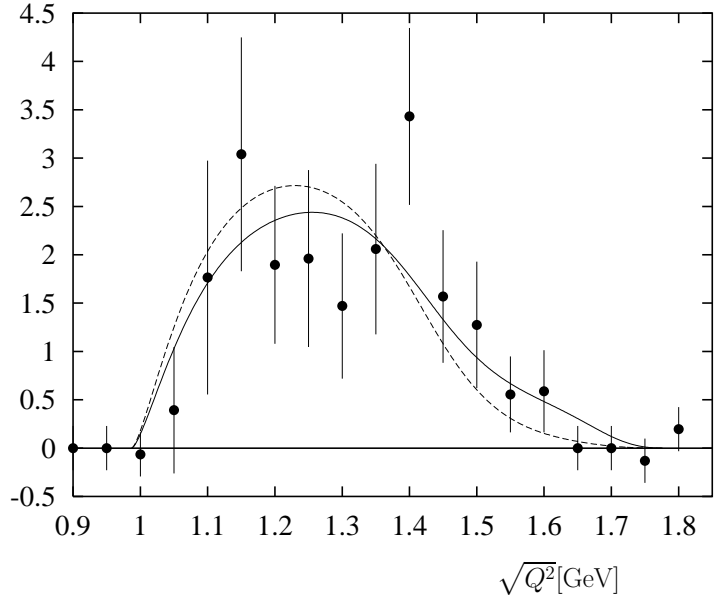}\label{fig-tau-a}
\caption{\it The normaized distribution 
$\frac{d\Gamma(\tau\to K^-K^0\nu_\tau)/d\sqrt{Q^2}}{
\Gamma(\tau\to K^-K^0\nu_\tau)}$ 
in the kaon pair invariant mass $\sqrt{Q^2}$ in units of $\mbox{GeV}^{-1}$
obtained from the fitted kaon form factor;
the solid (dashed) line corresponds to the constrained (unconstrained)
fit. The event distribution measured by CLEO Collaboration \cite{CLEOtau} 
and normalized, dividing by the total number of events, is shown with points. }
\end{figure}

\section{Conclusions}

In this paper we considered  the models of timelike form factors of pions
and kaons in \,anticipation of new and more accurate data
in the region above 1 GeV,  
from $e^+e^-$ machines using the radiative return method.
We introduced an ansatz for the pion form factor which is based 
on dual-resonance models and Veneziano amplitude. 
We argued that the parameters of the ground-state
and first excited states can deviate from the model
prediction due to effects of mixing with multiparticle (e.g., two -
or 4 $\pi$ states), therefore have to be fitted independently 
as free parameters. From the fit to the available pion form factor
data we have found that 
the main contribution to the form factor originates  from 
the ground state plus the first  two-three radially excited states.
The tail from the infinite series of resonances produces inessential,
but visible effects. The sign and value of the coefficients at certain 
excited resonances are shifted with respect to the dual 
QCD$_{N_c=\infty}$ ansatz signaling large mixing effects.  
Possible checks of the model are provided by the spacelike form factors,
the pion charge radius and the behaviour for large $s$ in the timelike region. 
In particular around $\sqrt{s}=3$ GeV we predict a value smaller 
than the one anticipated  from $J/\psi$ decay. 
On the other hand, data fitted to this model can be used for 
important tests of quark-hadron duality,
and of the QCD calculations of the pion form factor
in the spacelike region. 

Furthermore, we formulated an analogous model for the kaon form factor
and demonstrated that the contributions of 
$\phi$ , $\omega$ and $\rho$ resonances
(or, alternatively, the isospin 0 and 1 components) can be
separated by the fit. Interestingly, the $\tau\to K ^-K^0\nu_\tau$-decay 
distribution and partial width predicted from the model 
manifest a substantial sensitivity
to the pattern of $\rho$ resonances in the isospin-1 part of the form factor.

The model still has considerable room for improvement. In particular,
a more detailed kinematical and dynamical analysis
of total widths in the Breit-Wigner factors would allow 
to implement a more accurate energy-dependence in these widths.  

\bigskip

{\bf Acknowledgements}

We are grateful to S.~Eidelman for a useful discussion.
This work is supported by the 
German Ministry for Education and Research (BMBF).

\end{document}